\documentclass[12pt]{article}
\usepackage{amsmath}
\usepackage{amsfonts}
\usepackage{amssymb}
\usepackage{natbib}
\usepackage{enumitem}
\usepackage{url}
\usepackage[super]{nth}
\usepackage[utf8]{inputenc}
\usepackage[normalem]{ulem}
\usepackage{scalerel}
\usepackage{tabularx}
\usepackage{subfig}
\usepackage{graphicx}
\usepackage{subfig}
\usepackage{hyperref}
\usepackage{dsfont}
\usepackage{algorithm,algpseudocode}
\usepackage{blindtext}
\usepackage{setspace}
\usepackage[letterpaper, margin = 1in]{geometry}
\newcounter{algsubstate}
\renewcommand{\thealgsubstate}{\alph{algsubstate}}

\allowdisplaybreaks
\graphicspath{{}}

\title{Parameter estimation methods in Comparative Judgement}
\author{Ian Hamilton\footnote{Corresponding author: Ian Hamilton, Department of Statistics, University of Warwick, i.hamilton@warwick.ac.uk} , Nick Tawn}
\date{}

\begin{document}

\maketitle

\section*{Abstract}
Comparative Judgement is an assessment method where item ratings are estimated based on rankings of subsets of the items. These rankings are typically pairwise, with ratings taken to be the estimated parameters from fitting a Bradley-Terry model. Likelihood penalization is often employed. Adaptive scheduling of the comparisons can increase the efficiency of the assessment. We show that the most commonly used penalty is not the best-performing penalty under adaptive scheduling and can lead to substantial bias in parameter estimates. We demonstrate this using simulated and real data and provide a theoretical explanation for the relative performance of the penalties considered. Further, we propose a superior approach based on bootstrapping. It is shown to produce better parameter estimates for adaptive schedules and to be robust to variations in underlying strength distributions and initial penalization method.

%\keywords{adaptive testing; bias correction; pairwise comparison; penalized likelihood estimation}

\section{Background}\label{sn: background}

`Comparative judgement' (CJ) is a term used to describe a method of assessment by which a set of items are assessed based on rankings of subsets of the items via direct comparison. Most commonly the subsets consist of two items, with judges asked to provide a binary response indicating which item they consider to have more of some relevant quality, or simply to be better. The idea of using comparative judgement in this way builds on the idea that people are better at making comparative than absolute judgements (see, for example, \citet{laming2003human,goffin2011all}).

CJ was originally developed as a method in the social sciences \citep{bramley2007paired}, and has been proposed as both a summative and formative assessment mechanism in educational measurement (see, for example, \citet{pollitt2012method, jones2012summative, jones2017peer, davies2020comparative}). \citet{bartholomew2018systematic} and \citet{bartholomew2021systematized} provide systematic reviews of CJ in the context of K-16 education in the USA and higher education respectively, and \citep{wheadon2020comparative} is a study on what is currently the most widespread deployment of CJ in assessing primary school writing. In this work, we focus on its use in educational assessment but the method of pairwise comparison has been employed in other behavioral contexts such as health preference research [REF].

A topic of substantial investigation has been the method of scheduling the pairwise comparisons. Adaptive Comparative Judgement (ACJ) is a scheme for scheduling that has become popular and is embedded in one of the prominent CJ platforms, Digital Assess. It was originally proposed by one of the leading advocates of CJ, Alastair Pollitt, formerly of Cambridge Assessment, with a claimed increase in the efficiency of the assessments, achieving ratings of equivalent reliability from fewer judgements than with randomly scheduled comparisons \citep{pollitt2012comparative}. The method shares with other adaptive schemes the principle of scheduling comparisons between items of similar strength, such that each comparison yields more information. Other examples of such adaptive schemes in pairwise comparison include Swiss tournament scheduling and those proposed by \citet{revuelta1998comparison, glickman2005adaptive,  pfeiffer2012adaptive}. 

Data from the pairwise comparisons are typically analysed by considering the probability $p_{ij}$ that an item $i$ is preferred to an alternative $j$ in a direct comparison of the two items \citep{jones2023comparative}. The Bradley-Terry model \citep{bradley1952rank, zermelo1929berechnung} is used to analyse these data, where
\begin{equation}\label{eqn: glm BT definition}
    \text{Logit}(p_{ij}) = \lambda_i - \lambda_j,
\end{equation}
or alternatively expressed
\[
p_{ij} = \frac{\pi_i}{\pi_i + \pi_j},
\]
where $\lambda_i = \log(\pi_i)$ and will henceforth be referred to as the log-strength of $i$. 

The widespread adoption of the Bradley-Terry model in the CJ literature is perhaps due to its form as a dichotomous Rasch model \citep{andrich1978relationships}, with the family of Rasch models being familiar to educational researchers. More broadly, the Bradley-Terry model has statistical appeal, for example, in being the unique model for which the number of preferences (`wins') per item is a sufficient statistic \citep{buhlmann1963pairwise}.

An important consideration in assessments of this kind is the estimation approach. Maximum likelihood estimation is standard practice in fitting model parameters in CJ. But, in general, the CJ literature is frustratingly unclear on exactly what model penalizations, if any, are used and regrettably there does not appear to be a norm around making code and data available along with publication. The maximum likelihood log-strength estimates may be materially biased, especially when data is sparse. They will not even be finite if an item has been (dis-)preferred in all its comparisons. The large number of items and relatively small number of judgements in many CJ assessment exercises means these issues occur with high frequency. Using a penalization within the maximum likelihood estimation is a common way to achieve finiteness of estimates and reduce bias. Many studies use the No More Marking platform, which produces parameter estimates using a penalized estimation method \citep{wheadon2015analysing}.  It is under-appreciated that adaptive scheduling schemes such as ACJ and Swiss tournaments can dramatically accentuate the bias. Current penalization methods fail to mitigate for the extra complexity in these data generating structures that drive the bias. 

Bias in the strength parameters may effect the ordering of items and will bias standard measures of reliability to higher values. These reliability measures are employed to determine the number of comparisons required in an assessment, assess different scheduling schemes, and ultimately establish credibility for the CJ assessment method. Indeed, concerns about inflated reliability measures \citep{bramley2015investigating,bramley2019effect} have caused Digital Assess to reduce the adaptivity of its scheduling scheme \citep{rangel2018addressing}, and the largest CJ platform, No More Marking, to cease their investigations into using their own adaptive scheduling scheme \citep{wheadon2015opposite}. Adaptive scheduling schemes are more efficient, acquiring more information per judgement and therefore achieving the same reliability from fewer judgements. The reluctance to employ them because of issues with measuring reliability is thus holding back higher levels of efficiency in CJ assessment, which itself may be holding back CJ from wider adoption \citep{pinot2022classification}. An improved method of parameter estimation is therefore of substantial importance to the field.

The primary aim of this paper is to propose and demonstrate the value of alternative parameter estimation methods. In particular, we investigate penalizations not currently employed in the CJ literature and compare them to the current predominant method. A bootstrap method is also proposed and compared. The paper begins in Section \ref{sn: estimation} with a brief discussion of the literature on estimation methods and the identification and description of four penalties of interest. Section \ref{sn: simulation} compares these methods using a simulation study, taking into account varying underlying strength distributions and scheduling schemes. It shows that one of these methods seems to notably outperform the others for the adaptive scheduling scheme tested. Section \ref{sn: empirical} augments this simulation-based analysis with analysis of real data, allowing an analysis on non-simulated data and an alternative adaptive scheduling scheme. Section \ref{sn: inference} presents a theoretical explanation for the relative performance of the estimation methods observed in the previous sections. Section \ref{sn: bias correction} proposes a bootstrap method for bias correction and investigates the effectiveness of this approach by means of the simulation study. It shows that the bias can be substantially eliminated and that the method shows good robustness to underlying strength distribution, scheduling scheme and original estimation method combinations. Section \ref{sn: concluding} offers some concluding remarks, summarising the implications for future practice and research.

\section{Estimation methods}\label{sn: estimation}
Various methods of estimation have been proposed for contexts relevant to CJ. For example, one stream of literature considers bias reduction in relevant model families \citep{firth1993bias, kosmidis2009bias, kosmidis2011multinomial, kosmidis2014bias, kenne2017median, kosmidis2020mean, kosmidis2021jeffreys}. Another, independent, stream of investigation considers estimation in the context of Rasch models, typically considering both bias and predictive accuracy \citep{molenaar1995estimation, linacre2004rasch, haberman2004joint, robitzsch2021comprehensive}. A third stream considers Bayesian estimation and the selection of an appropriate prior within a Bradley-Terry model \citep{davidson1973bayesian, leonard1977alternative, chen1984bayes, whelan2017prior, phelan2017hierarchical}. In this section, those literatures as well as the CJ literature itself are leant on in selecting four estimation methods to review. We investigate the most commonly used penalized estimation approach in the CJ literature, as well as three others that have particular appeal, but appear not to have been used in CJ. 

All the methods considered in this section are forms of penalized likelihood estimation, a method based on adding a penalty function into the score equation. Prior to penalization, the score equation under the Bradley-Terry model for an item $r$ is 
\[
\frac{\partial l(\boldsymbol{\lambda})}{\partial \lambda_r} = w_r - \sum_{j} m_{rj}p_{rj},
\]
where $l(\boldsymbol{\lambda})$ is the log-likelihood, $w_r = \sum_j c_{rj}$ is the number of observed preferences for item $r$ over all other items, with $c_{ij}$ the number of times $i$ has been preferred to $j$ during the CJ assessment, and $m_{rj}$ is the number of comparisons between $r$ and $j$. Under maximum likelihood estimation, the score function is set to zero and the model parameters $\boldsymbol{\lambda}$ estimated from this system of equations.

Under a penalized likelihood approach, a penalty term, $a_r$, is introduced into the score equation, 
\[
w_r + a_r = \sum_{j \neq r} m_{rj}p_{rj} \quad \text{for all }r.
\]
This has the effect of ensuring the finiteness of estimates even where items have been (dis)preferred in all comparisons, and of providing shrinkage to the estimates, such that bias may be reduced. The first penalization method appears to be the predominant one in the CJ literature.

\subsection{\texorpdfstring{$\epsilon$}{epsilon}-adjustment \texorpdfstring{\citep{bertoli2014estimating}}{(Bertoli-Barsotti et al 2014)}}
Many CJ studies have been performed using the No More Marking platform, which is made freely available to researchers. For example, a Google Scholar search for ``comparative judgement" OR ``comparative judgment'' AND ``nomoremarking'' yields 127 results as at 26\textsuperscript{th} March 2024. \citet{wheadon2015analysing} indicates that No More Marking use the \textbf{btm} function from the \textbf{sirt} package \citep{robitzsch2022package} in \textbf{R} \citep{RCore2021alanguage} for their analysis. This function uses a bias reduction method proposed by \citet{bertoli2014estimating}, which they call the $\epsilon$-adjustment approach, which takes a penalty
\begin{equation}\label{eqn: epsilon}
a_r = \epsilon \left(1 - 2\frac{w_r}{m_r}\right), 
\end{equation}
where $m_r = \sum_j m_{rj} = \sum_j \left( c_{rj} + c_{jr} \right)$ is the number of comparisons involving $r$, and $\epsilon$ is a configurable constant, set to be $0.3$ by default in \textbf{sirt}. The penalty for item $r$ is thus related to its observed win ratio. The number of observed preferences, a sufficient statistic for the log-strengths under the Bradley-Terry model, is therefore transformed from being in the interval $[0,m_r]$ to $[\epsilon, m_r-\epsilon]$ for each $r$, ensuring that log-strength estimates are finite. \citet{robitzsch2021comprehensive} found the $\epsilon$-adjustment to be one of the best-performing of a wide variety of proposed estimation methods when considering the wider context of Rasch models, looking at bias and root mean squared error. None of the other estimation methods discussed here was considered in that study, though the method of \citet{warm1989weighted}, which was considered and rejected, is closely related to that of \citet{firth1993bias}.

\subsection{\texorpdfstring{$\alpha$}{alpha}-adjustment}\label{sn: alpha}
\citet{davidson1973bayesian} propose a Bayesian approach with a conjugate $\text{Beta}(\gamma, \gamma)$ prior on each of the pairwise preference probabilities. A natural choice is to take $\gamma = \alpha / (n-1)+1$, since this implies a score penalty of
\begin{equation}\label{eqn: alpha}
a_r =  \alpha \left( 1 - 2\frac{\sum_{j \neq r} p_{rj}}{n-1}   \right),
\end{equation}
so that the penalty can be seen to relate to the average probability of item $r$ being preferred to all other items. It is equivalent to using a penalty whereby each item is assumed to have been preferred to each other item an additional $\alpha / (n-1)$ times. \citet{crompvoets2020adaptive} uses a similar penalty, but without the $1/(n-1)$ scaling, which has the undesirable feature that the penalty becomes stronger with more items.

Using the $\alpha$-adjustment penalty, the likelihood is augmented by a multiplicative term that is a function of the pairwise information,
\[
\prod_{i,j} p_{ij}^{\alpha/(n-1)} =
\prod_{i<j}(p_{ij}(1-p_{ij}))^{\alpha/(n-1)} = \prod_{i<j}(I_{ij})^{\alpha/(n-1)}.
\]
We will later show the relevance of this information-based framing in Section \ref{sn: inference}.  .

\subsection{Dummy item \texorpdfstring{\citep{phelan2017hierarchical}}{(Phelan 2017)}}
\citet{phelan2017hierarchical} propose a method that takes a dummy item of average quality, against which each item is preferred and dispreferred an equal number of times, $c_0$, which need not be an integer. This method seems to have been used in early implementations of the Bradley-Terry model applied to rank teams in college hockey in the USA with $c_0$ taken to be a half \citep{wobus2007KRACH}. As \citet{phelan2017hierarchical} note, this is equivalent to taking a Bayesian approach where the prior is a $\text{Beta}(c_0+1, c_0+1)$ distribution on the probability of an item of zero log-strength being preferred to each item $i$. In the `dummy item' approach the likelihood is therefore augmented with a multiplicative term
\[
\prod_{i} p_{i0}^{c_0}(1-p_{i0})^{c_0},
\]
where $0$ denotes the dummy item, and $p_{i0}$ the probability of item $i$ being preferred to the dummy item. The log-strength of the dummy item is set to zero, which also provides an identifiability constraint.

The dummy item adjustment provides a score penalty of
\begin{equation}\label{eqn: dummy}
a_r = c_0\left(1 - \frac{2e^{\lambda_r}}{1 + e^{\lambda_r}}\right) = c_0\left(1 - 2p_{r0}\right) ,
\end{equation}
so where the $\epsilon$- and $\alpha$-adjustments depended on win ratio and average preference probability, the dummy item penalty depends on the probability of being preferred to the (dummy) item of mean quality.

\subsection{Firth (1993)}\label{sn: Firth}
The most cited work on the topic of bias reduction of maximum likelihood estimates is \citet{firth1993bias}. It notes that previous proposals to use a jackknife method \citep{quenouille1949approximate, quenouille1956notes} or `corrective' bias correction rely on estimates being finite, but this cannot be guaranteed in general. Instead a `preventive' correction is proposed, based on eliminating the O(1/n) bias term (see, for example, \citet[p.455-456]{mccullagh1989generalized}). In the context of exponential family distributions, such as the Bradley-Terry model, this may be described by an adjustment to the log-likelihood by the log of the square-root of the determinant of the information matrix, $i(\boldsymbol{\lambda})$,
\[
l^*(\boldsymbol{\lambda}) = l(\boldsymbol{\lambda}) + \frac{1}{2} \log \begin{vmatrix} i(\boldsymbol{\lambda}) \end{vmatrix},
\]
which is alternatively viewed as the \citet{jeffreys1946invariant} invariant prior for the problem. Note that here we have not specified if this is observed or expected information, since for the Bradley-Terry model, conditional on the comparisons observed being an ancillary statistic, the observed and expected information matrices are equal. In contrast, under an adaptive scheduling scheme the comparisons observed are no longer an ancillary statistic, but we do not generally have means of computing the expected information, so for this exercise we use the observed information. The implications of this will be discussed at more length in Section \ref{sn: inference}.

In terms of the score function, the penalization proposed is an additive term to the number of preferences $w_r$ of 
\begin{equation}\label{eqn: Firth}
a_r = \frac{1}{2} \text{tr}\left\{i(\boldsymbol{\lambda})^{-1}\left(\frac{\partial i(\boldsymbol{\lambda})}{\partial \lambda_r}\right)\right\}.
\end{equation}

\citet{firth1993bias} observes that the method is equivalent to solutions of the maximum likelihood equations on adjusted data of $c^*_{ij} = c_{ij} + h_{ij}/2$ and $m^*_{ij} = m_{ij} + h_{ij}$, where $h_{ij}$ is the leverage of comparisons between items $i$ and $j$, a measure of how far away the observation of pair $i,j$ is from other observations, derived from the hat matrix. The `adjusted data' score equation is thus
\[
w^*_r = \sum_{j \neq r} m^*_{rj}p_{rj}
\]
where
\[
w^*_r = \sum_{j \neq r} c^*_{rj} = \sum_{j \neq r} \left( c_{rj} + \frac{h_{rj}}{2} \right) = w_r + \frac{1}{2}\sum_{j \neq r} h_{rj}.
\]
The score equation being solved in the maximum likelihood estimation would then be
\[
w_r + \frac{1}{2}\sum_{j \neq r} h_{rj} = \sum_{j \neq r} p_{rj}\left(m_{rj} + h_{rj}\right).
\]
Rearranging, the penalty term may then be expressed as 
\begin{equation}\label{eqn: Firth generalized form}
 a_r = \frac{1}{2}\sum_{j \neq r} h_{rj} \left(1 - 2 \frac{\sum_{j \neq r} p_{rj}h_{rj}}{\sum_{j \neq r} h_{rj}} \right),   
\end{equation}
so that the penalty for item $r$ can be seen to relate to a leverage-weighted average probability of it being preferred to all other items.

\section{Simulation study}\label{sn: simulation}
\subsection{Study design}\label{sn: sim design}
In order to investigate the predominant $\epsilon$-adjustment penalization method in comparison to the alternatives, we first use a simulation study. The distribution of true underlying log-strengths of any item set is unknown. Indeed, this is what we are looking to make inference on. It is highly desirable to have estimation methods that are robust to the heterogeneity of these underlying distributions. Therefore, we use three distributions with distinct characteristics --- normal, bimodal and skew normal. The standard deviation of the log-strengths is taken to be 2, in line with the findings of \citet{rangel2018addressing}, and the mean to be zero. The three distinct distributions are used to generate three separate collections of log-strengths from which data from a Bradley-Terry data-generating process is simulated. For each distribution, 100 items are considered. This is in line with the order of magnitude of many assessments based on a university or school cohort. 

\begin{figure}
    \centering
    \includegraphics[width=14cm]{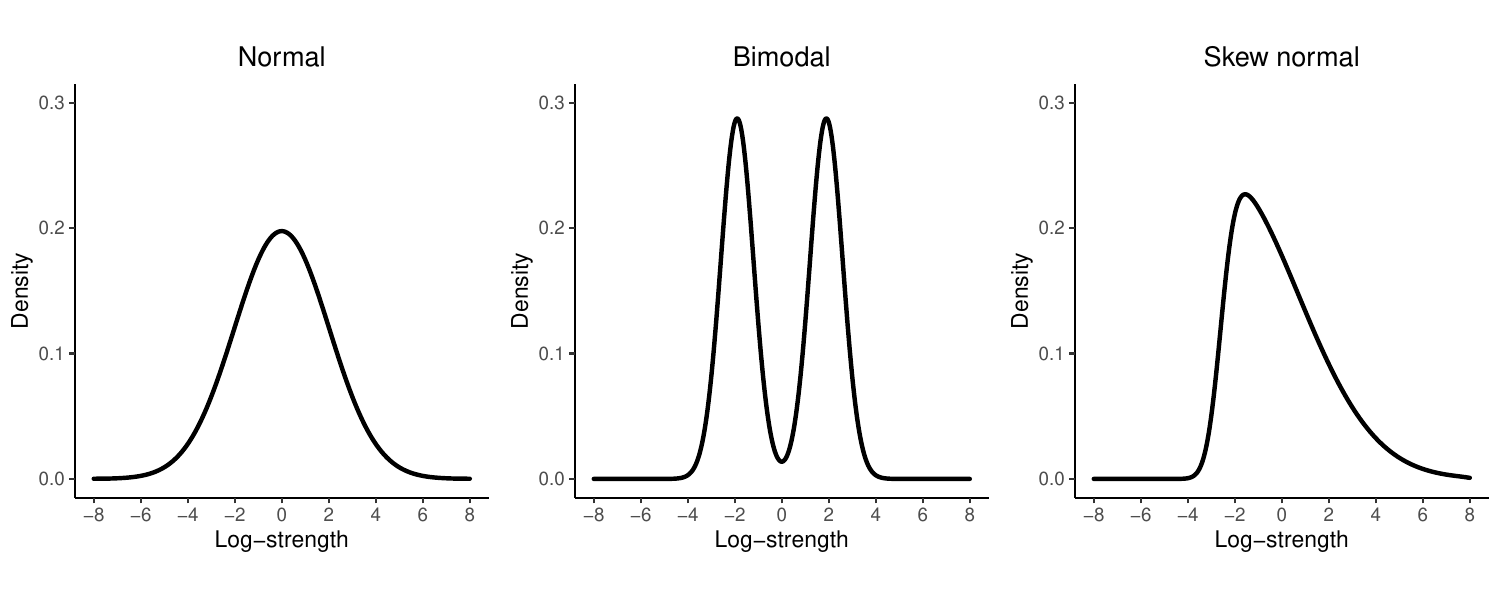}
    \caption{Densities for the underlying log-strength distribution for simulation}
    \label{fig: densities}
\end{figure}

Thus, for the three distributions, the log-strength of the $k$th item is taken to be ($k=1, \dots ,100)$:
\begin{enumerate}
    \item {\makebox[3cm][l]{\text{Normal: }} $2\Phi^{-1}\left((k-0.5)/100 \right)$}
    \item {\makebox[3cm][l]{\text{Bimodal: }} $\frac{2}{3.174}\left( \Phi^{-1}\left((k-0.5)/50 \right) - 3 \right), k= 1, \dots, 50 ;$} \\
    {\makebox[3.5cm][l]{} $\frac{2}{3.174}\left( \Phi^{-1}\left((k-50.5)/50 \right) + 3 \right), k= 51, \dots, 100 $}
    \item {\makebox[3cm][l]{\text{Skew normal: }} $2\Psi^{-1}\left((k-0.5)/100; \alpha=8, \omega = 3.274, \xi =2.592  \right)$},
\end{enumerate}
where $\Phi$ is the cumulative distribution function for a standard normal distribution, and $\Psi$ is the cumulative distribution function for a skew normal distribution with $\Psi_X(x; \alpha, \omega, \xi) = \Phi((x- \xi)/\omega) - 2T((x- \xi)/\omega,\alpha)$ where $T(h,a)$ is Owen's T function. These give the densities shown in Figure \ref{fig: densities}

For the dummy item method, $c_0$ will be taken to be 0.25. In testing based on a round-robin tournament, where the expected penalty was calculated analytically, this value was found to provide a good approximation to the method of \citet{firth1993bias}. For the $\epsilon$-adjustment approach, $\epsilon$ will be taken to be 0.3. \citet{robitzsch2021comprehensive} suggests that in the wider context of a Rasch model a choice of $\epsilon=0.24$ is optimal, but here we are interested in what practitioners might do and it seems not unreasonable to consider that many would take the default setting when applying the function. In order to be able to make a more direct comparison to the $\epsilon$-adjustment, and given the comparability of their structures as expressed in equations (\ref{eqn: epsilon}) and (\ref{eqn: alpha}), $\alpha$ is also taken to be 0.3. For the purpose of this investigation, the assessments will be taken to consist of 20 rounds of comparison with each item appearing in one comparison in each round. The choice of 20 is based on personal correspondence with No More Marking where they confirmed that this was the standard request they made of submitting schools. That is, if a school submits 100 items to be assessed then the school would be expected to perform a thousand pairwise comparisons, so that each item is compared 20 times.

Two scheduling schemes will be investigated. Both will consist of rounds of comparison where each item is compared once in each round. In the first scheduling scheme, the pairs in each round are selected uniformly at random. In the second scheduling scheme, the pairs are selected according to a Swiss system. Under the Swiss scheduling scheme, the first round of pairings is random. In each subsequent assessment round, items are paired with other items with as similar as possible (typically the same) number of preferences up to that point. 

While the Swiss system is not formally used in CJ, it forms the basis for the ACJ method, being used in the first four rounds of that scheme and having the same underlying philosophy of matching items of similar strength. For the purposes of this investigation it allows for the approaches to be assessed against an intuitive scheme that is computationally cheaper to simulate, not requiring the re-estimation of strength parameters after each round that ACJ and other schemes do, and avoids the complication of determining an appropriate method for the intermediate parameter strength estimation. 

Thus, given the three underlying distributions and the two scheduling schemes, six distribution and scheduling scheme combinations were simulated using a Bradley-Terry data generating process. 1000 assessments were simulated under each of the combinations. The results were then analysed using each of the four different estimation methods. All simulations are performed in \textbf{R} \citep{RCore2021alanguage}. The \citet{firth1993bias} penalization is fitted using the \textbf{brglm2} package \citep{kosmidis2020brglm2}, the other penalizations are fitted using a Gauss-Siedel algorithm based on the code used in the \textbf{btm} function in \textbf{sirt} \citep{robitzsch2022package}. 

Based on these simulations, estimates of the bias and mean absolute error for each item are calculated as
\begin{equation}\label{eqn: bias}
    \text{Bias}_i = \frac{1}{N}\sum^{N}_{k=1} \lambda_{ik} - \lambda^*_i,
\end{equation}
\begin{equation}\label{eqn: mean absolute error}
    \text{Mean absolute error}_i = \frac{1}{N}\sum^{N}_{k=1} \mid \lambda_{ik} - \lambda^*_i \mid,
\end{equation}
where $\lambda_{ik}$ is the estimate for the log-strength of the i\textsuperscript{th} item $(i = 1, \dots , 100)$ from the k\textsuperscript{th} simulation $(k = 1, \dots , N)$, with $N=1000$ in this case, and $\lambda^*_{i}$ is the `true' log-strength of item $i$ used to generate the simulation.

\begin{figure}
    \centering
    \includegraphics[width=13cm]{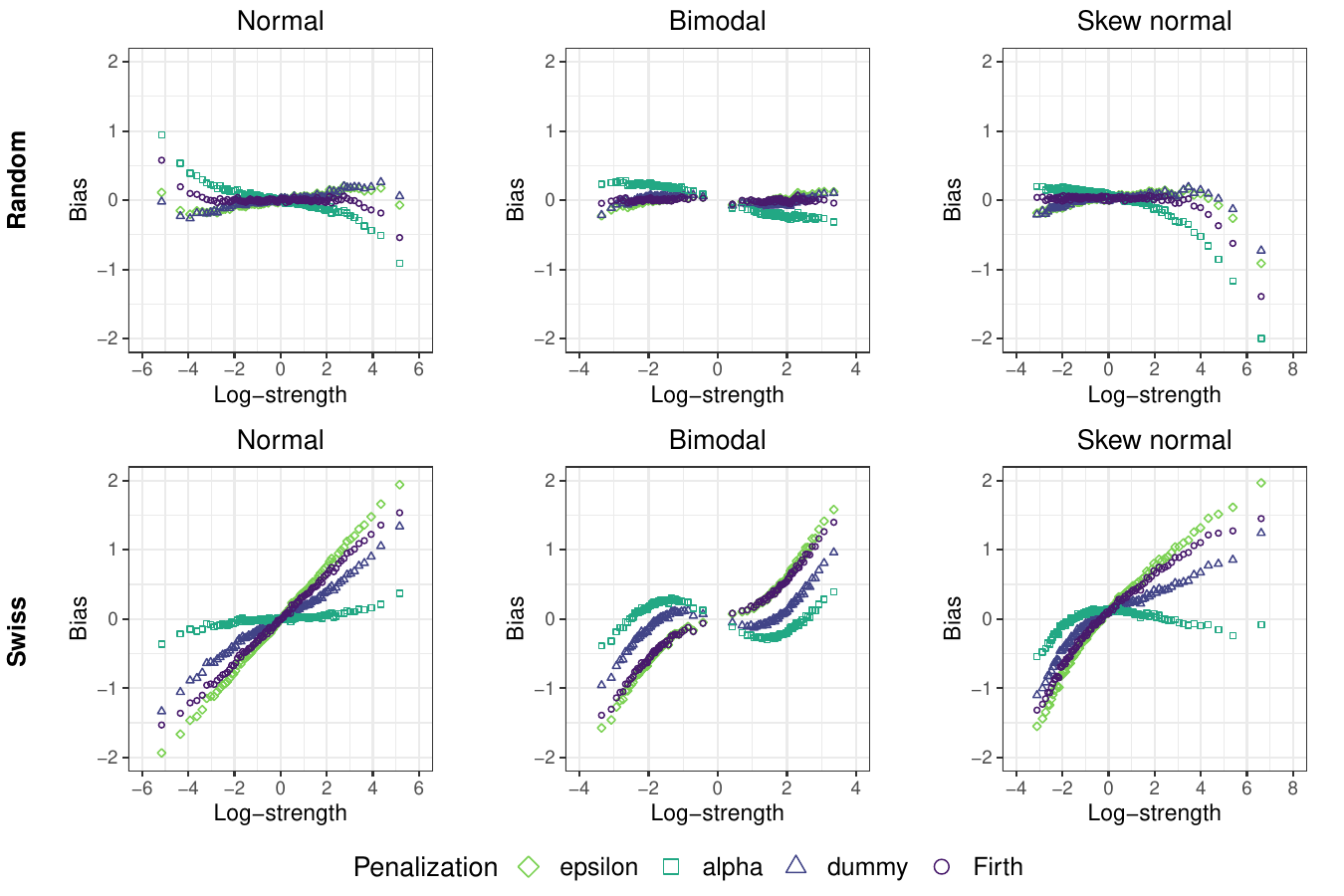}
    \caption{Bias of log-strength estimates under different scheduling scheme, log-strength distribution and penalization method combinations. All four methods achieve substantially reduced bias for random schedules, but only $\alpha$-adjustment is effective in reducing bias under a Swiss scheme.}
    \label{fig: bias}
\end{figure}

\begin{figure}
    \centering
    \includegraphics[width=13cm]{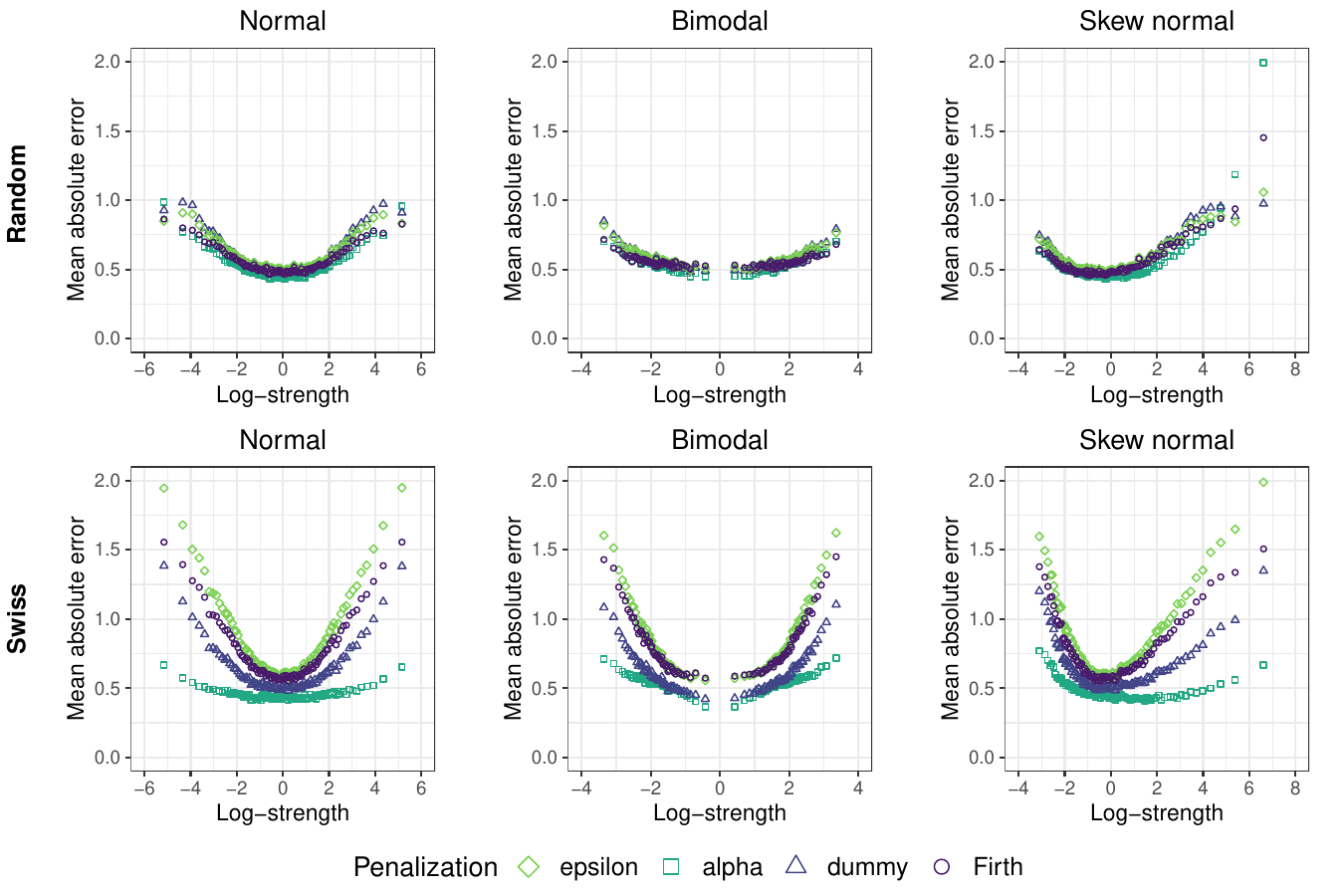}
    \caption{Mean absolute error of log-strength estimates under different scheduling scheme, log-strength distribution and penalization method combinations. All four methods achieve substantially reduced error for random schedules, but only $\alpha$-adjustment is effective in reducing error under a Swiss scheme.}
    \label{fig: EAE}
\end{figure}

\begin{figure}
    \centering
    \includegraphics[width=13cm]{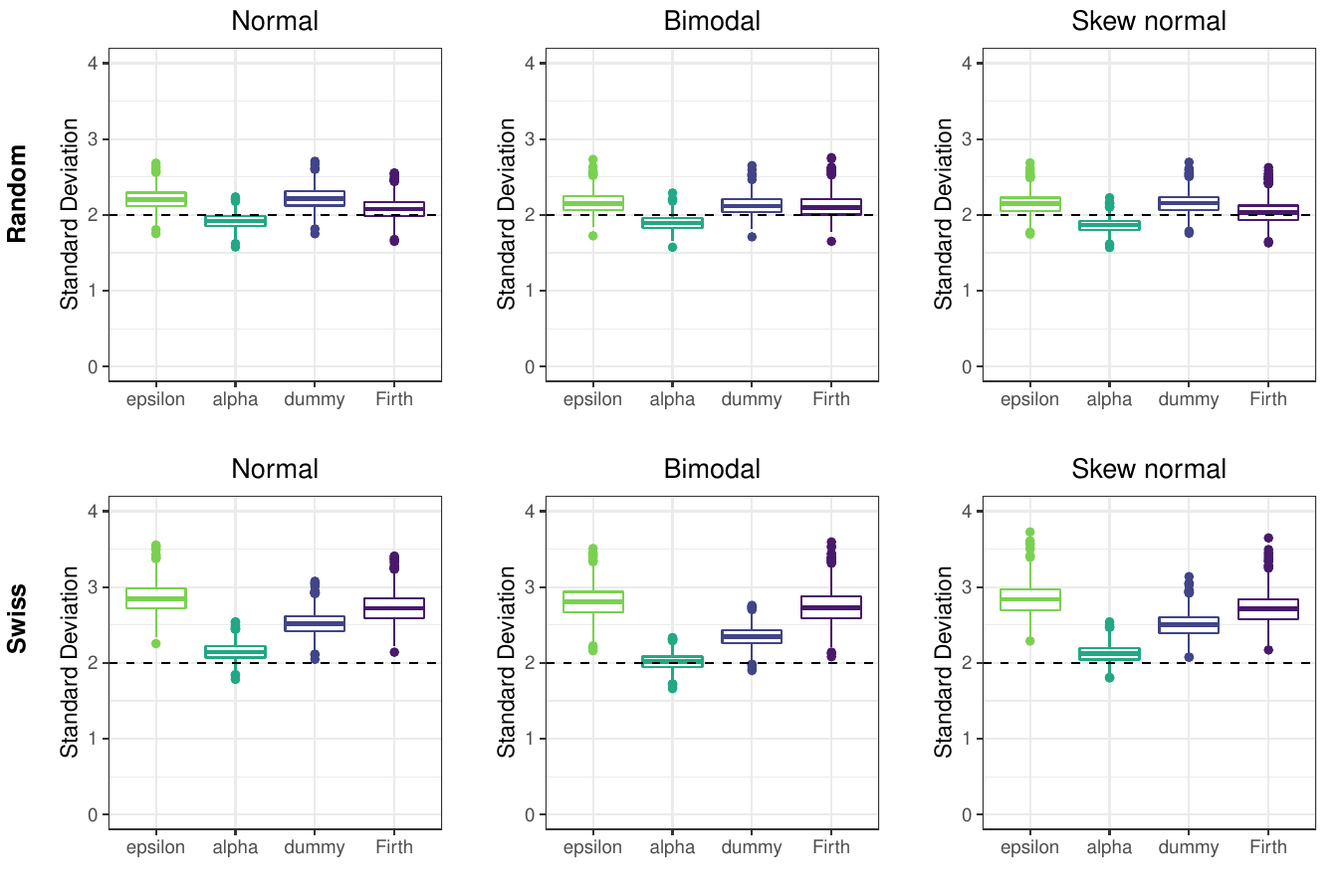}
    \caption{Boxplots illustrating the sampling distribution of the estimator for the standard deviation of the log-strengths under both the random and Swiss scheduling schemes. In the case of Swiss tournaments, only $\alpha$-adjustment achieves a good estimate of the true standard deviation of 2, represented by the dotted line.}
    \label{fig: sd}
\end{figure}

\subsection{Results}\label{sn: sim results}
For the randomly generated schedule, Figures \ref{fig: bias} and \ref{fig: sd} suggest that all four estimation methods do an effective job in constraining the bias and reducing error. 

Under the Swiss scheduling scheme there is more divergence between the estimation methods. Figure \ref{fig: bias} shows that there is an extremising tendency in the estimation of the log-strengths for all the methods, where the lower strength item estimates have a negative bias and the higher strength item estimates have a positive bias. However, the $\alpha$-adjustment gives notably lower absolute error and bias than the other methods. Figure \ref{fig: sd} confirms these findings and shows that any remaining extremisation under the $\alpha$-adjustment is limited to a very small number of items as the overall standard deviation is in line with that of the generating distribution, in contrast to the other penalization methods. This suggests that the $\alpha$-adjustment might be a better choice than the alternatives considered here. However, the wider applicability of the conclusion might be challenged on the grounds of the data set being simulated rather than real, and the adaptive scheduling being a Swiss tournament, rather than a more sophisticated algorithm. Thus, in the next section, we look at a real data set that uses an alternative, more sophisticated, scheduling algorithm.

\section{Comparison of methodologies on real data}\label{sn: empirical}
In order to complement the simulation study of the previous section, this section seeks to investigate parameter estimation using a real data set. For this purpose, the results from \citet{bramley2019effect} are reanalysed. \citet{bramley2019effect} was a follow-up study to \citet{bramley2015investigating}, which used a simulation-based study to demonstrate that adaptive scheduling schemes could have an inflationary effect on the Scale Separation Reliability (SSR), the most commonly used reliability measure in CJ. That work was criticised for the simulations used not being representative of real world data \citep{pollitt2015reliability}. In response, \citet{bramley2019effect} sought to collect relevant real data.

The data consist of pairwise comparisons from three CJ assessment exercises. The assessment exercises are referred to here, consistently with \citet{bramley2019effect}, as studies 1a, 1b, and 2. They were designed as follows:
\begin{itemize}
    \item[] 1a --- a study of 150 GCSE English essays, with comparisons scheduled by an adaptive scheme, made by 18 judges.
    \item[] 1b --- a study of a subset of 20 GCSE English essays from the wider set of 150, with a round-robin format where every one of the 20 items was compared with every other item once. The judges were the same as those in 1a.
    \item[] 2 --- a study of the same 150 GCSE English essays, with comparisons scheduled randomly. There were 16 judges, none of whom had participated in 1a and 1b. In addition, three pairs of items were selected based on their rating from study 1a and these pairs were compared multiple times. These additional comparisons have been removed in the analysis of study 2 that follows, so that it remains representative of a randomly scheduled assessment. They have been included in the combined results, so that those include as much data as possible. 
\end{itemize}

The judges were all examiners of English GCSE for the Oxford, Cambridge and RSA (OCR) examining body. In study 1b, 19 judges were recruited to do 10 judgements each. However, one judge dropped out late in the process and the results of another were excluded based on their poor consistency with other judges and their response time (average of 1 second per judgement). This means that each item is compared to just 17 rather than 19 others in study 1b. A summary of the data is given in Table \ref{tbl: Bramley data}. For further details of data collection, see \citet{bramley2019effect}. 

\begin{table}[htbp!]
\centering
\begin{tabular}{c|cccc}

            & 1a adaptive   & 1b round-robin   & 2 random    & \shortstack{Combined \\ (1a, 1b, 2)}  \\
\hline
Essays     & 150   & 20    & 150   & 150 \\
Judges     & 18    & 17    & 16    & 34 \\
\shortstack{Judgements \\ per essay}    
            & 14.4  & 17    & 13.3  & 31.3      
\end{tabular}
\caption{Data summary for studies from \citet{bramley2019effect}}
\label{tbl: Bramley data}
\end{table}

The adaptive scheme used in study 1a was based on the progressive method proposed in \citet{revuelta1998comparison} and further discussed in \citet{barrada2008incorporating, barrada2010method}. This method has intuitive appeal as it provides for a gradual transition from scheduling pairs on a random basis to scheduling pairs with the most information. This information is calculated based on intermediate estimations of item strength. Unfortunately, the specifications of the implementation of the scheduling algorithm are not available, precluding the application of our novel bootstrap method introduced later in this paper.

We use the data here in three ways. First, the assessment of the same items through an adaptive and a random scheme enables us to examine the interaction between estimation method and scheduling scheme in parameter estimation. Second, the three studies taken together give a large amount of data per item, so that we can benchmark the standard deviation of the estimates based on studies 1a and 2 alone to those based on the combined data. Third, when the data from all studies are combined, the 20 items in study 1b have been judged a mean of 45 times, including a direct comparison between each other in almost all cases. We therefore take the log-strength estimates for these items from the combined data as a `quasi-true' log-strength, against which the estimations from studies 1a and 2 are compared.

\citet{bramley2019effect} states that maximum likelihood estimation was used to calculate the log-strength estimates under a Bradley-Terry model. It does not discuss the use of a penalty, and indeed the results presented in Table \ref{tbl: Bramley standard deviation}, showing the standard deviation of the log-strength estimates, strongly suggests that none was used. For study 2, there are eight essays which were either preferred or dispreferred in all comparisons. It is reported that these ``receieved a measure based on an extrapolation rule'', though this is not specified. 

Results in Table \ref{tbl: Bramley standard deviation} are strongly suggestive of bias in the estimates reported by \citet{bramley2019effect}, due to not using an adequate penalty, especially in the case of the adaptive scheme. This led to the unintuitive finding that the preferred estimate of reliability in CJ assessment, SSR, was higher based on the data from study 1a alone than when analysing the data combined over all studies. 

We can also use the data in Table \ref{tbl: Bramley standard deviation} to investigate the relative performance of the four estimation methods. The combined data has a large number of judgements per essay and so we might expect the standard deviation of the log-strength estimates to be consistent across different estimation methods and indicative of a true range. We do indeed see much more consistency to the standard deviation of log-strength estimates using the combined data, with a range of 1.24-1.45 logits across the estimation methods. When we compare this value to the standard deviation based on the adaptive study 1a alone we see that of the estimation methods tested, the $\alpha$-adjustment comes closest, but even that requires a stronger parametrisation of $\alpha=0.5$ to constrain the standard deviation to within that range. It is perhaps to be expected that a greater scaling term would be required under this scheme than under the Swiss scheduling scheme, as the Swiss scheme is less sophisticated in identifying comparators of a close strength and thus less adaptive. Taking $\alpha=0.5$ seems to over-constrain in the case of the random scheduling scheme, which might give us cause to be cautious of advocating its universal use.

\begin{table}[htbp!]
\centering
\begin{tabular}{c|ccc}

            & 1a adaptive   & 2 random    & \shortstack{Combined \\ (1a, 1b, 2)}  \\
\hline
\citet{bramley2019effect}   & 4.65   & 1.77    & 1.52 \\
No penalty     & 4.60    & N/A    & 1.52  \\
\citet{firth1993bias}     & 4.02    & 1.39    & 1.40  \\
$\epsilon$-adjustment ($\epsilon = 0.3$)     & 3.76    & 1.56    & 1.45  \\
Dummy ($c_0=0.25$)     & 2.59    & 1.56    & 1.43  \\
$\alpha$-adjustment ($\alpha = 0.3$)    
            & 1.91  & 1.32    & 1.34   \\
$\alpha$-adjustment ($\alpha = 0.5$)    
            & 1.43  & 1.15    & 1.24   
\end{tabular}
\caption{Estimated log-strength standard deviation as reported in \citet{bramley2019effect} and using no penalty, \citet{firth1993bias} and $\alpha$-adjustment.}
\label{tbl: Bramley standard deviation}
\end{table}

\begin{figure}[hbtp]
    \centering
    \includegraphics[width=14cm]{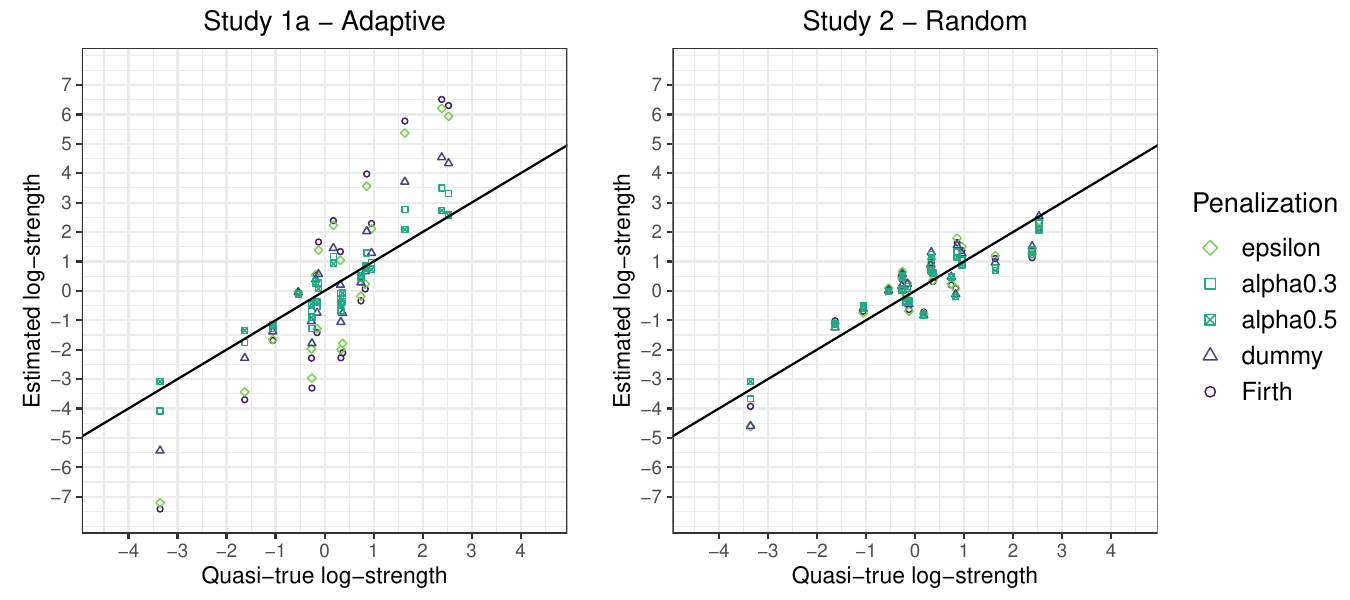}
    \caption{Log-strength estimates for the 20 items in study 1b based on comparisons from studies 1a and 2 of \citet{bramley2019effect} plotted against quasi-true values based on the combined data sets from studies 1a, 1b, 2.}
    \label{fig: Bramley 1a 2 lambda}
\end{figure}

For the estimation of the `quasi-true' log-strengths using the combined data we use $\alpha$-adjustment with $\alpha=0.3$, though the conclusions are not sensitive to this selection compared to using the other penalizations. Figure \ref{fig: Bramley 1a 2 lambda} shows that the penalization methods perform similarly under the random scheduling scheme in study 2, with only the `No penalty' approach deviating substantially. However, under the adaptive scheme of study 1a, the deviation is much greater and the $\alpha$-adjustment achieves estimates much closer to the `quasi-true' log-strengths than the other methods. For this subset of items, $\alpha=0.5$ performs better than $\alpha=0.3$ in estimating the items at the extremes of the strength distribution but very similarly for other items.

It may be possible to address the bias under the $\epsilon$-adjustment by using a higher value of $\epsilon$, but doing so would meaningfully over-constrain in the case of random scheduling and would cause it to deviate further from the value used by default in the \textbf{sirt} package \citep{robitzsch2022package} and the value \citet{robitzsch2021comprehensive} found to be optimal. This demonstrates that the appropriate value of $\epsilon$ is a function of the level of adaptivity of the scheduling scheme, but this will likely not not be known beforehand as it will be dependent on the underlying strength distribution. In the next section, we demonstrate that there are reasons to believe that the outperformance of the $\alpha$-adjustment was not just due to a fortunately chosen parametrisation, but rather relates to the nature of the penalty.

\section{Theoretical underpinnings of bias in CJ}\label{sn: inference}
In the previous sections, we demonstrated that there is greater bias to estimates under an adaptive scheduling scheme than under a random scheduling scheme, and that different penalizations are differently able to mitigate this. 

In this section, we provide some theoretical understanding for why this is the case. By decomposing the likelihood, we show that the parameter estimates under the two scheduling schemes will be the same given the same data. However, we go on to explain how the estimators under adaptive scheduling schemes will have greater bias. This motivates the consideration of a pseudo-likelihood that explains the stronger performance of the $\alpha$-adjustment approach. 

Consistent with elsewhere in this paper, an `observed comparison' refers to two items being compared and does not include the outcome of that comparison.  The outcome will be referred to as a `preference'. Introducing some additional notation, let 
\begin{itemize}
    \item $_rX_{ij},  (r \in \{1, \dots , R\})$, be a random variable that takes value 1 if $i$ is preferred to $j$ in round $r$ and 0 otherwise.
    \item ${}_rx_{ij}$ be the observed preference from the $r$th round of an assessment for a pair of items $(i,j)$, where ${}_rx_{ij}$ is 1 if $i$ has been preferred to $j$ in round $r$ and 0 otherwise.
    \item $\boldsymbol{x} = ({}_1x_{12}, {}_1x_{13}, \dots , {}_1x_{(n-1)n}, {}_2x_{12}, \dots , {}_2x_{(n-1)n}, \dots ,{}_Rx_{12}, \dots , {}_Rx_{(n-1)n})$ be the observed sample of preferences.
    \item ${}_rY_{ij}$ be a random variable that takes value 1 if $i$ is compared to $j$ in round $r$ and 0 otherwise, so that ${}_rY_{ij} = {}_rY_{ji} = {}_rX_{ij} + {}_rX_{ji}$.
    \item ${}_ry_{ij}$ be the observed comparison from the $r$th round of the assessment for a pair of items $(i,j)$, where ${}_ry_{ij}$ is 1 if $i$ has been compared to $j$ in round $r$ and 0 otherwise, so that ${}_ry_{ij} = {}_ry_{ji} = {}_rx_{ij} + {}_rx_{ji}$.
    \item $\boldsymbol{y} = ({}_1y_{12}, {}_1y_{13}, \dots , {}_1y_{(n-1)n}, {}_2y_{12}, \dots , {}_2y_{(n-1)n}, \dots ,{}_Ry_{12}, \dots , {}_Ry_{(n-1)n})$ be the observed sample of comparisons.
    \item ${}_rC_{ij} = \{{}_rX_{ij} = {}_rx_{ij}\}$ be the event that the preference between $i$ and $j$ in round $r$ was as observed.
    \item $C_r = \cap_{i,j}{}_rC_{ij}$ be the event that the preferences in round $r$ were as observed.
    \item $C = \cap_{r} C_r = \cap_{r}\cap_{i,j} \thickspace {}_rC_{ij}$ be the event that the preferences during each round of the assessment were as observed.
    \item ${}_rM_{ij} = \{{}_rY_{ij} = {}_ry_{ij}\}$ be the event that the comparison between $i$ and $j$ in round $r$ was as observed.
    \item $M_r = \cap_{i,j} {}_rM_{ij}$ be the event that the comparisons in round $r$ were as observed.
    \item $M = \cap_{r} M_r = \cap_{r}\cap_{i,j} \thickspace {}_rM_{ij}$ be the event that the comparisons during each round of the assessment were as observed.
    \item $m_{ij} = \sum^R_{r=1} {}_ry_{ij}$ be the total number of observed comparisons between $i$ and $j$ over the whole assessment.
\end{itemize}
Additionally, to recap notation introduced earlier,
\begin{itemize}
    \item $p_{ij}$ is the probability that $i$ is preferred to $j$ given a comparison between $i$ and $j$.
    \item $c_{ij} = \sum^R_{r=1} {}_rx_{ij}$ is the total number of observed preferences for $i$ over $j$ during the assessment.
    \item $\boldsymbol{\lambda}$ is the vector of log-strengths of the items, where we assume a Bradley-Terry data generating process, such that $p_{ij} = \mathbb{P}({}_rX_{ij} =1 | {}_rY_{ij} = 1) = e^{\lambda_i} / (e^{\lambda_i} + e^{\lambda_j})$.
\end{itemize}

\subsection{Decomposing the likelihood}
Under a maximum likelihood approach, in the present setting, we observe a CJ assessment and wish to estimate the log-strengths, $\boldsymbol{\lambda}$. We are considering the likelihood $L (\boldsymbol{\lambda}; \boldsymbol{x})$, which, for a tournament of $R$ rounds has the same form as
\[
 p(\boldsymbol{x};\boldsymbol{\lambda})= \mathbb{P}(C;\boldsymbol{\lambda})= \mathbb{P}(C_R, \dots , C_1; \boldsymbol{\lambda}).
\]
Conditioning on the observed comparisons in round $R$, and noting that conditional on the event $M_R$ and given $\boldsymbol{\lambda}$, $C_R$ is independent of ${C_{R-1}, \dots , C_1}$ by the assumption of our data-generating process and that conditional on ${C_{R-1}, \dots , C_1}$, $M_R$ has no dependence on $\boldsymbol{\lambda}$,
\begin{align*}
    \mathbb{P}(C;\boldsymbol{\lambda}) &= \mathbb{P}(C_R \mid M_R, C_{R-1}, \dots , C_1; \boldsymbol{\lambda}) \mathbb{P}(M_R \mid C_{R-1}, \dots , C_1; \boldsymbol{\lambda}) 
    \mathbb{P}(C_{R-1}, \dots , C_1; \boldsymbol{\lambda}) \\
    &= \mathbb{P}(C_R \mid M_R ; \boldsymbol{\lambda}) \mathbb{P}(M_R \mid C_{R-1}, \dots , C_1) 
    \mathbb{P}(C_{R-1}, \dots , C_1; \boldsymbol{\lambda}).
\end{align*}
Applying the same reasoning iteratively to $\mathbb{P}(C_{R-1}, \dots , C_1,; \boldsymbol{\lambda})$ we have that
\begin{equation*}
  \mathbb{P}(C, M ; \boldsymbol{\lambda}) = \prod^R_{r=1} \mathbb{P}(C_r \mid M_r ; \boldsymbol{\lambda}) \prod^R_{r=2} \mathbb{P}(M_r \mid C_{r-1}, \dots , C_1) \mathbb{P}(M_1).
\end{equation*}
Under the assumptions of our data-generating process the preferences within each round are independent of each other and so
\[
\prod^R_{r=1} \mathbb{P}(C_r \mid M_r ; \boldsymbol{\lambda}) = \prod^R_{r=1} \prod_{i,j} \mathbb{P}({}_rC_{ij} \mid {}_rM_{ij} ; \boldsymbol{\lambda}) = \prod_{i,j} \prod^R_{r=1} p_{ij}^{{}_rx_{ij}} = \prod_{i,j} p_{ij}^{c_{ij}},
\]
giving
\begin{equation}\label{eqn: non-conditional}
\mathbb{P}(C ; \boldsymbol{\lambda}) =  \prod_{i,j} p_{ij}^{c_{ij}} \prod^n_{r=2} \mathbb{P}(M_r \mid C_{r-1}, \dots , C_1) \mathbb{P}(M_1).
\end{equation}

Note that the term $\prod^n_{r=2} \mathbb{P}(M_r \mid C_{r-1}, \dots , C_1)\mathbb{P}(M_1)$ is not dependent on $\boldsymbol{\lambda}$, and so the estimation of $\boldsymbol{\lambda}$ is entirely dependent on $\prod_{i,j} p_{ij}^{c_{ij}}$, whose form is dictated by the conditional independence assumption of the Bradley-Terry model, whether the data have been generated under a random or adaptive scheme. Thus, given the same data, the estimate will be the same whether generated under an adaptive or random scheduling scheme. 

However, the estimators will not have the same sampling distribution under the random and adaptive schemes. This is because the comparisons observed under an adaptive scheme are more likely to be between two items closer in strength. For example, if two strong items, close in strength, are compared, one must be preferred and its strength will then be estimated to be very high since it was preferred to another strong item. With enough comparisons, the relative strengths of these two items would be accurately reflected in their proportion of wins, but under finite (and often sparse) sampling, adaptive scheduling schemes are more likely to produce sets of preferences where individual log-strengths are extremised in this way and hence the estimator has greater bias. The estimates under an adaptive or random scheduling scheme will be the same given the same preferences, but preferences that produce more extreme strength parameters are more likely to occur under adaptive scheduling and thus the scheduling schemes produce estimators with different bias while applying the same estimation method.

An analogous example of this type of problem is provided in \citet{barlow2020inference}. There, data is generated according to a Negative Binomial with a set number of successes $X$ and probability of success $p$. \citet{barlow2020inference} identify that the maximum likelihood estimate of the probability of a success is the same whether the data was generated by a Binomial or Negative Binomial distribution, provided that the observed data had the same number of trials and successes. Thus, in both cases, $\hat{p} = X/N$, where $X$ is the number of successes and $N$ is the number of trials. However, the estimator properties are very different. In the Binomial case the estimate is unbiased, whilst in the Negative Binomial case the estimate is biased. In the CJ setting, this issue is less transparent due to the lack of an analytic expression for the estimator.

\subsection{Conditional likelihood}
One approach to address this bias, is to consider conditioning on the total observed comparison set in the inference,
\begin{equation*}
    \mathbb{P}(C; \boldsymbol{\lambda}) = \mathbb{P}(C \mid M; \boldsymbol{\lambda})\mathbb{P}(M; \boldsymbol{\lambda}). 
\end{equation*}

Under a random scheduling scheme,  given $\boldsymbol{\lambda}$, $C_r$ is dependent only on $M_r$. The observed comparisons are independent of other rounds and of $\boldsymbol{\lambda}$, so that $\mathbb{P}(M ; \boldsymbol{\lambda}) = \mathbb{P}(M)$ is uniform --- each possible tournament has equal probability of being observed. Alternatively expressed, $\{m_{ij} : i,j \in \{1, \dots , n\}, i \neq j\}$ is an ancillary statistic, and it should be conditioned on for inference \citep{cox1958some}. 

Under an adaptive scheme, the observed comparisons are dependent on the item-strengths. While the game selections for the first random round of adaptive tournaments is an ancillary statistic, the later rounds where adaptive scheduling occurs are not. Under schemes that seek to maximise information on a pairwise basis, items that are close in strength are more likely to be compared. Relatedly, it is no longer the case that $C_r$ is generally independent of other round comparisons conditional on $M_r$, since $M_{r+1}$ will have a dependency on $C_k$ for $k \leq r$.

However, it is not clear how the terms in this likelihood formulation, $\mathbb{P}(C \mid M ; \boldsymbol{\lambda})$ and $\mathbb{P}(M ; \boldsymbol{\lambda})$, may be modelled even knowing the scheduling scheme. For this reason, we might consider instead a pseudo-likelihood, which conditions on the observed comparisons for each round, where we take an appropriate scheduling distribution for each round that reflects the strengths. 
\begin{equation}\label{eqn: pseudo-likelihood}
\mathbb{P}(C; \boldsymbol{\lambda}) \approx \prod^R_{r=1} \mathbb{P}(C_r , M_r ; \boldsymbol{\lambda}) = \prod^R_{r=1} \mathbb{P}(C_r \mid M_r ; \boldsymbol{\lambda})\mathbb{P}(M_r ; \boldsymbol{\lambda}).
\end{equation}
Under the assumption of a Bradley-Terry data-generating process,
\begin{equation*}\label{eqn: B-T likelihood}
\prod^R_{r=1} \mathbb{P}(C_r \mid M_r ; \boldsymbol{\lambda}) = \prod_{i,j} p^{c_{ij}}.
\end{equation*}
Based on equation (\ref{eqn: pseudo-likelihood}), we might therefore expect a penalty function to perform better, the closer it approximates $\prod^R_{r=1} \mathbb{P}(M_r ; \boldsymbol{\lambda})$.

In any round, $r$, the number of comparisons is constrained such that each item is compared once. Thus, $\prod^R_{r=1} \mathbb{P}(M_r ; \boldsymbol{\lambda})$ is positively related to 
\[
\prod^R_{r=1} \prod_{(i,j) \in \mathcal{M}_r} \mathbb{P}({}_rY_{ij} = 1 ; \boldsymbol{\lambda}),
\]
where $\mathcal{M}_r = \{(i,j) : {}_ry_{ij} = 1 \}$ is the set of pairs $(i,j)$ that were compared in round $r$. Under an adaptive scheme, the probability $\mathbb{P}({}_rY_{ij} = 1 ; \boldsymbol{\lambda})$ will be higher when the items are closer in strength. It is this insight that is suggestive of why the $\alpha$-adjustment seems to work better than alternatives.

\begin{figure}[hbtp]
    \centering
    \includegraphics[width=14cm]{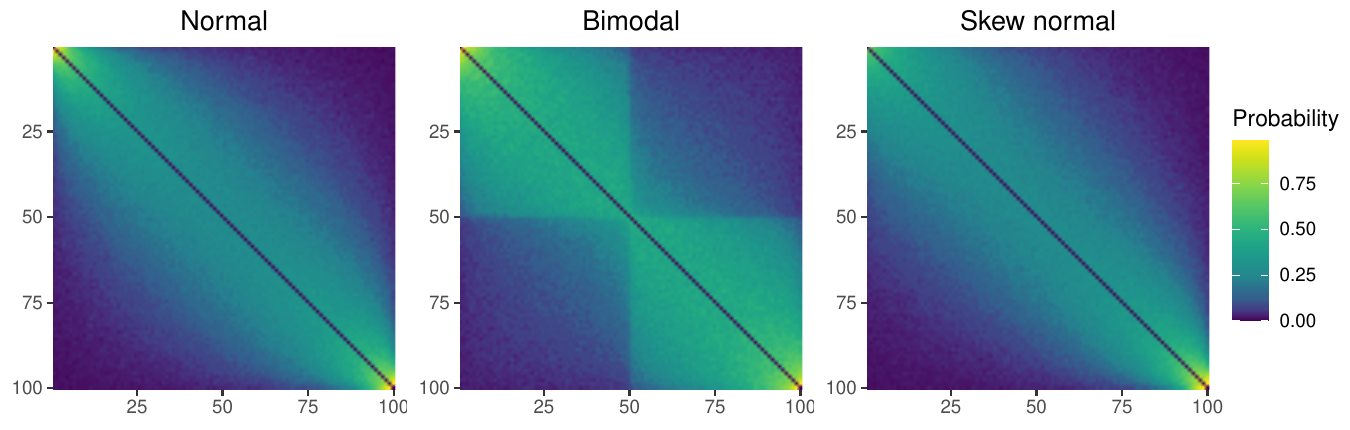}
    \caption{Probability of at least one comparison for each pair of items under a 20 round Swiss scheduling scheme, based on 1000 simulated assessments. Items shown in strength order from 1, the weakest to 100, the strongest.}
    \label{fig: probability of comparison}
\end{figure}

Here, this is illustrated graphically by computing the average probability of observing at least one comparison between pair $(i,j)$ over the twenty comparison rounds of the Swiss scheduling scheme,
\[
\mathbb{P}\left( \sum^{20}_{r=1}{}_rY_{ij} \geq 1 ; \boldsymbol{\lambda}\right).
\]
To calculate this, we use the simulations from Section \ref{sn: simulation} and for each pair $(i,j)$ calculate the proportion of simulations where at least one comparison was observed. The calculation is made for each of the three log-strength distributions. Results are shown in Figure \ref{fig: probability of comparison}.

\begin{figure}[hbtp]
    \centering
    \includegraphics[width=14cm]{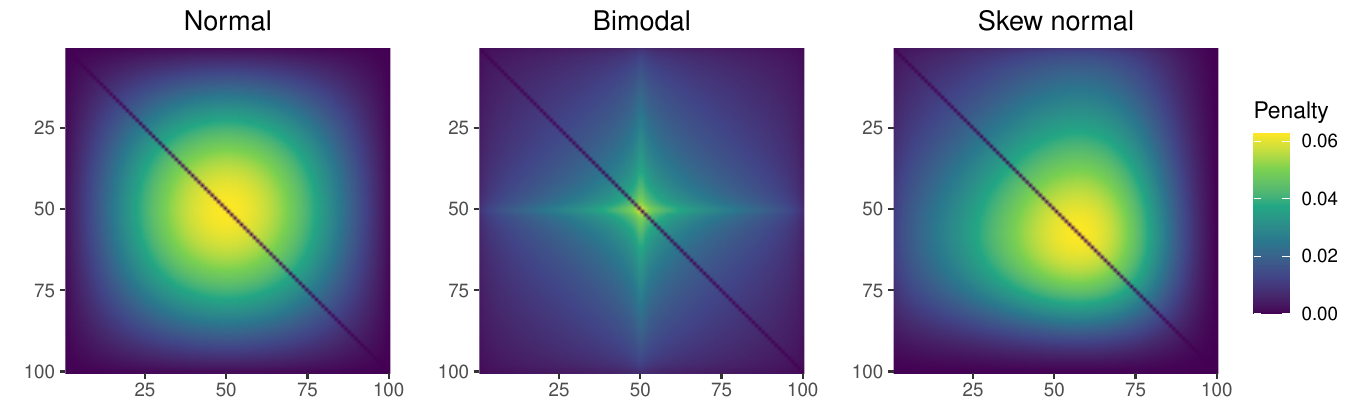}
    \caption{Pairwise value of dummy penalty. Items shown in strength order from 1, the weakest to 100, the strongest.}
    \label{fig: dummy prior}
    \centering
    \includegraphics[width=14cm]{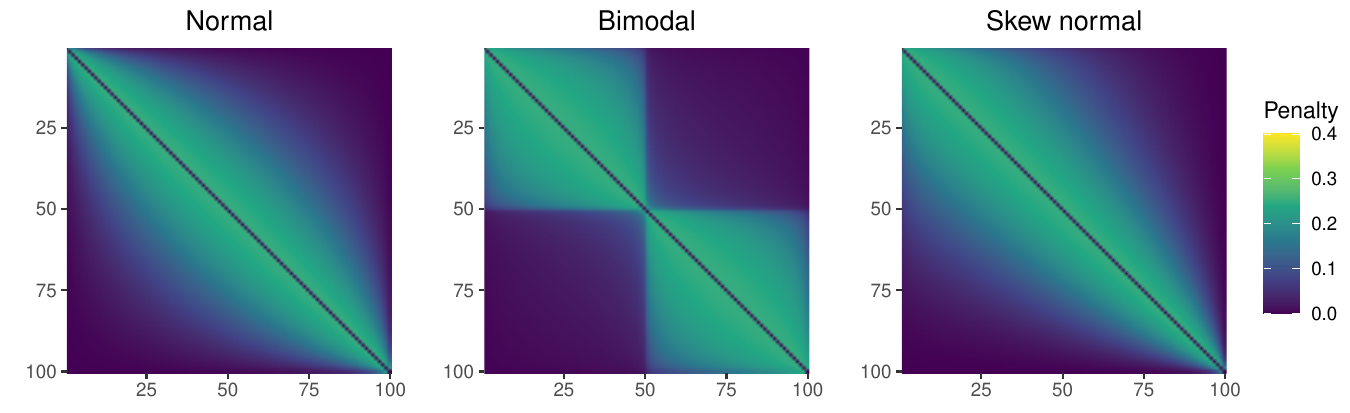}
    \caption{Pairwise value of $\alpha$-adjustment penalty. Items shown in strength order from 1, the weakest to 100, the strongest.}
    \label{fig: alpha prior}
\end{figure}

Recall that under the dummy penalty approach, we consider a likelihood penalty of 
\[
\prod_i p_{i0}^{c_{0}}(1-p_{i0})^{c_{0}} = \prod_{i < j} (p_{i0}p_{0i}p_{j0}p_{0j})^{c_0 / (n-1)},
\]
where the dummy item represented by $0$ has log-strength of zero. Under the $\alpha$-adjustment, we consider a likelihood penalty of
\[
\prod_{i,j} p_{ij}^{\alpha / (n-1)} = \prod_{i< j} (p_{ij}p_{ji})^{\alpha / (n-1)} .
\]
Thus, these likelihood penalties can be decomposed in terms of a value relating to pair $(i,j)$ and compared to the average probability of observing a comparison between that pair. $p_{i0}p_{0i}p_{j0}p_{0j}$ and $p_{ij}p_{ji}$ respectively are plotted for the three log-strength distributions in Figures \ref{fig: dummy prior} and \ref{fig: alpha prior}.

The intensity of the profiles for these penalties is adjustable by varying the parameters $c_0$ and $\alpha$ respectively, but the shape of the profile will remain the same. There is a clear consistency of profile between the $\alpha$-adjustment in Figure \ref{fig: alpha prior} and the probability of observing at least one comparison in Figure \ref{fig: probability of comparison}. On the other hand, the profiles diverge much more for the dummy item adjustment seen in Figure \ref{fig: dummy prior}. This is suggestive of why the $\alpha$-adjustment, with an appropriately selected value for $\alpha$, seemed to provide better estimates in Sections \ref{sn: simulation} and \ref{sn: empirical} when using the adaptive scheme, since it shows graphically how the penalty that the $\alpha$-adjustment provides is consistent with the term $\prod_r \mathbb{P}(M_r ; \boldsymbol{\lambda})$ from the pseudo-likelihood in equation (\ref{eqn: pseudo-likelihood}), in line with the observation in Section \ref{sn: alpha} that the $\alpha$-adjustment may be interpreted as an information-based penalty. We might also note that under a random scheduling scheme the intensity profile would be flat. The profile of the $\alpha$-adjustment penalty deviates from a flat profile more than the dummy item adjustment. This is consistent with the observation that the $\alpha$-adjustment showed slightly greater bias under a random sampling scheme in Figure \ref{fig: bias}. 

The other two penalties, $\epsilon$-adjustment and \citet{firth1993bias}, do not give pairwise likelihood penalties that allow us to compare them in the same way. However, there are independent reasons to expect that they may not be performant for adaptive scheduling schemes. The $\epsilon$-adjustment relies on the ratio $w_r / m_r$ increasing with item strength. But this will be the case to a far lesser degree under an adaptive scheme, where comparisons are more likely between items of similar strength. Under a random scheduling scheme, the $w_r / m_r$ term will approximate the $\sum_j p_{rj} / (n-1)$ term of the $\alpha$-adjustment. Under an adaptive scheduling scheme, this results in too little bias correction when using the $\epsilon$-adjustment as seen in Figure \ref{fig: bias}. It is possible that the penalty would perform better with a greater value of $\epsilon$, but it seems likely to be highly sensitive to the adaptivity of the scheduling scheme, the number of rounds of comparison and the distribution of the item strengths, which may not all be well anticipated prior to analysis. It therefore seems likely to be an inferior penalty to the $\alpha$-adjustment for use with adaptive scheduling schemes. The \citet{firth1993bias} penalty applies an adjustment to the likelihood based on the information matrix. Under random sampling, the schedule is an ancillary statistic and so should be conditioned on in calculating the information matrix \citep{cox1958some}. Under the Bradley-Terry model, this leads the observed and expected information matrices to be equivalent. Under an adaptive scheduling scheme, the schedule is no longer an ancillary statistic and the observed and expected information matrices are not equal. The expected information matrix is not generally known for any given adaptive scheduling scheme. Here, in line with how practitioners may use available software in analysis of these data, we used the observed information matrix as an alternative, but this cannot be expected to capture the asymptotic bias of the estimator. 

This section has established theoretical grounds for the $\alpha$-adjustment's superiority as a penalty for use with adaptive schemes. This complements the results observed in Sections \ref{sn: simulation} and \ref{sn: empirical}. However, the $\alpha$-adjustment failed to eliminate bias and the apparent requirement to use a larger $\alpha$ for the real data results using the more adaptive scheme discussed in Section \ref{sn: empirical} indicates there is no single optimal value for $\alpha$ applicable to all scheduling scheme and strength distribution combinations that we can expect to more satisfactorily eliminate bias. This can therefore act as encouragement to investigate an alternative approach.

\section{Bias-corrected estimation}\label{sn: bias correction}
In this section, we consider the use of a bootstrap method for correcting the bias of the log-strength estimates. Bootstrapping is a well established approach for bias correction. Albeit computationally heavy, bootstrapping is particularly useful when an explicit preventative penalization is intractable. We have such an example here with adaptive scheduling. In Section \ref{sn: inference} we showed that only the choice of first round comparisons are ancillary (in contrast to all game choices in a random scheduling). As such, to correctly apply the method of \citet{firth1993bias} in line with \citet{cox1958some} then one should use the conditional information. This is intractable without solving something combinatorially hard. The other penalization methods are invariant to the tournament structure.

We introduce a bias correction bootstrapping procedure for adaptively scheduled tournaments. This method correctly and fully conditions on the appropriate ancilliary statistic. Indeed, the implications and relevance of this method for bias correction in an adaptively scheduled setting extend beyond the education literature. The bias correction method here works first by estimating the log-strengths of a Bradley-Terry model based on the results of the CJ assessment of interest. These log-strength estimates are then used to simulate other CJ assessments using the same scheduling scheme and a Bradley-Terry data-generating process. Crucially, in resimulating the assessments, we maintain the part of the scheduling that is an ancillary statistic. For the randomly scheduled tournament that is the entire tournament. For the Swiss scheme that is just the first round. The log-strengths are then estimated for each of these simulated CJ assessments. The bias of the original estimate is calculated as the mean of the difference in the simulated item log-strength estimates of the simulations and the original estimate. This bias is then subtracted from the original log-strength estimate to get a bias-corrected estimate.
That is, the bias-corrected item log-strength estimates are
\begin{equation}\label{eqn: bias correction}
    {}_{BC}\lambda_i = {}_0 \lambda_i - \left( \frac{1}{m}\sum^m_{s=1}  {}_s\lambda_i - {}_0 \lambda_i \right),
\end{equation}
where ${}_0 \lambda_i$ is the original estimate for the log-strength of item $i$, and ${}_s\lambda_i, (s = 1, \dots, m)$ is the log-strength estimate for item $i$ from simulation $s$ within the bootstrap calculation. 

\begin{figure}[hbtp]
    \centering
    \includegraphics[width=14cm]{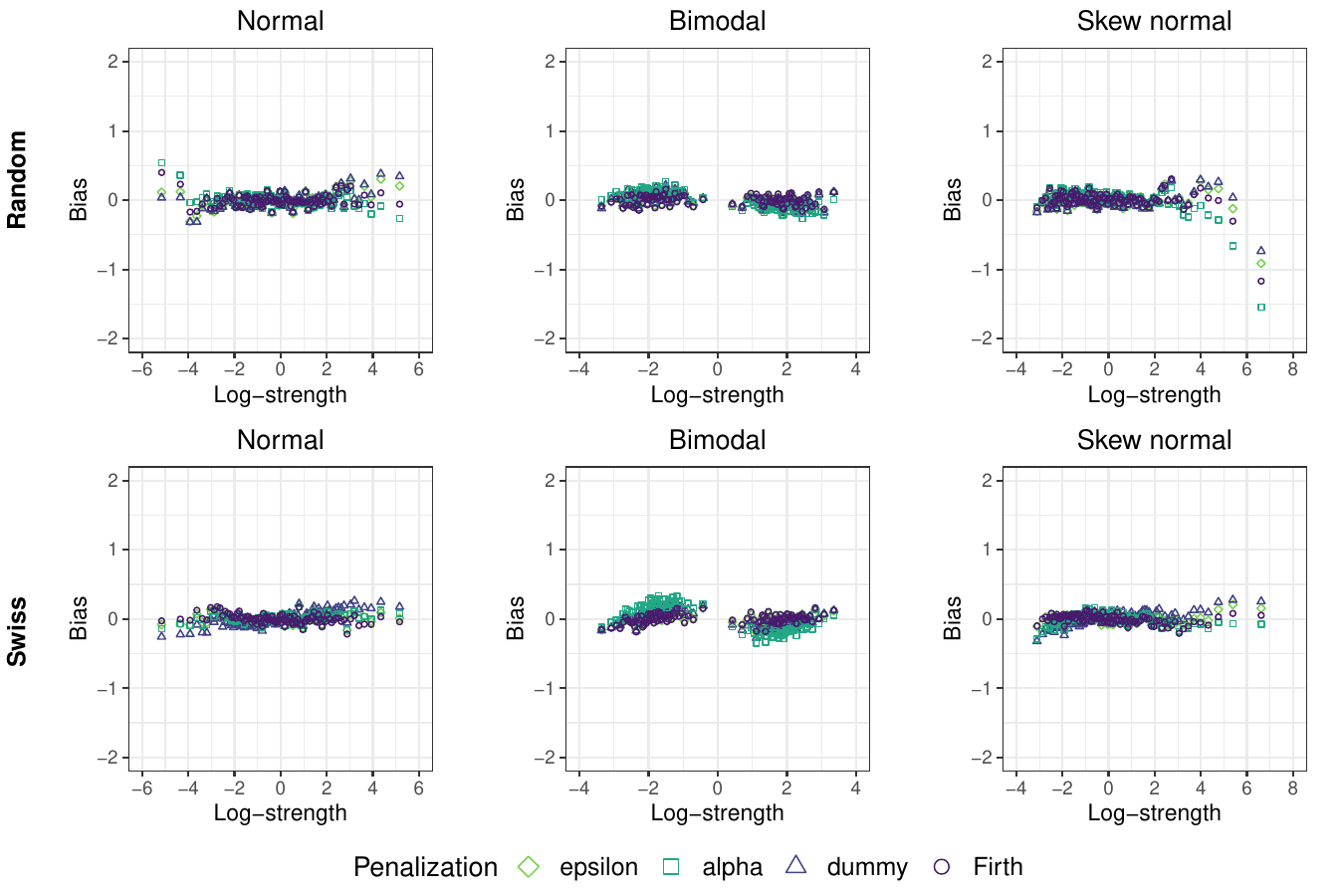}
    \caption{Bias using bias-corrected estimator under different scheduling scheme, log-strength distribution and penalization method combinations. Bias is reduced to close to zero in all cases.}
    \label{fig: bias bias-corrected}
\end{figure}

\begin{figure}[hbtp]
    \centering
    \includegraphics[width=14cm]{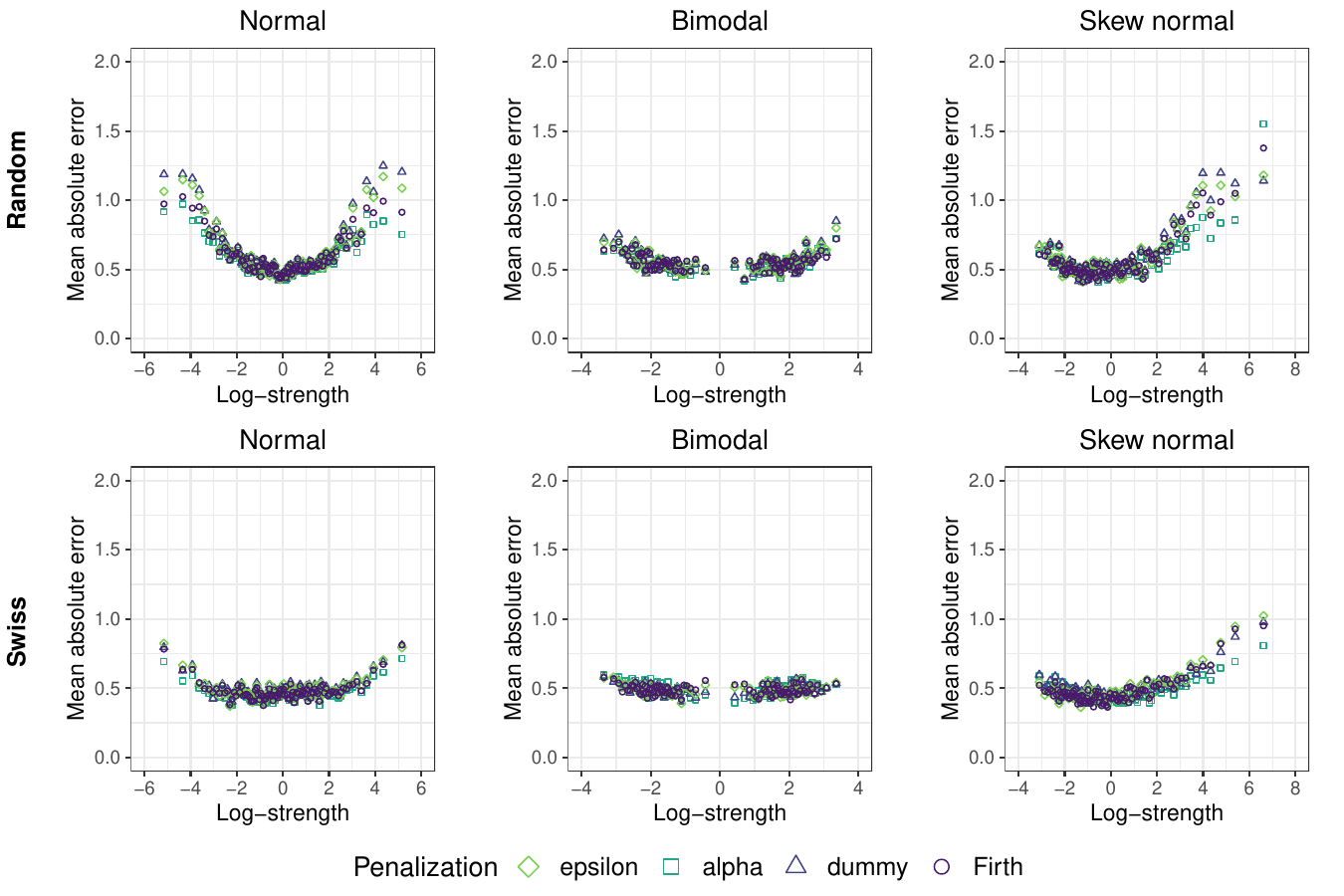}
    \caption{Mean absolute error using bias-corrected estimator under different scheduling scheme, log-strength distribution and penalization method combinations.}
    \label{fig: EAE bias-corrected}
\end{figure}

The bias correction procedure is summarised in Algorithm \ref{alg: bias-correct sim}. In the algorithm we utilise the distinction between a tournament, which we defined as a schedule of comparisons not including the outcomes of those comparisons, and an assessment, which is a schedule of comparisons including their associated outcomes. We test the bias-corrected estimates through extending the simulation study. We take $m=40$ in Algorithm \ref{alg: bias-correct sim} and simulate 100 assessments for each of the six scheduling and strength distribution combinations. As before, each tournament has 100 items and 20 rounds. Having estimated these bias-corrected log-strengths, $\boldsymbol{\lambda}^{BC}$, we calculate the estimated bias and mean absolute error as defined in equations (\ref{eqn: bias}) and (\ref{eqn: mean absolute error}). The results are presented in Figures \ref{fig: bias bias-corrected} and \ref{fig: EAE bias-corrected}. 

\begin{algorithm}
\caption{bias correction algorithm}\label{alg: bias-correct sim}
\begin{algorithmic}[1]
\Require{$T,A$,m}.
\State Estimate log-strengths, $\boldsymbol{\lambda}$, based on judgements from assessment $A$. 
\If{Scheduling = random} 
    \State Simulate assessment $A_s$ from tournament $T$ using a Bradley-Terry data-generating process and log-strengths $\boldsymbol{\lambda}$.
\Else
\If{Scheduling = Swiss} 
    \State Simulate assessment $A_s$ using only first round of tournament $T$ and later rounds based on Swiss scheduling using a Bradley-Terry data-generating process and log-strengths $\boldsymbol{\lambda}$.
\EndIf
\EndIf
\State Estimate log-strengths, $\boldsymbol{\lambda}^s$, based on judgements from assessment $A_s$.
\State Repeat steps 2-9 $m$ times.
\State Calculate bias-corrected log-strength estimates $\boldsymbol{\lambda}^{BC}$ using equation (\ref{eqn: bias correction}).
\end{algorithmic}
\end{algorithm}

For the random schedule, a comparison to Figures \ref{fig: bias} and \ref{fig: EAE} shows that the bias correction offers little performance advantage over penalization. However for the Swiss scheme, both the bias and mean absolute error are substantially reduced. The results also suggest that bias correction is robust to the underlying strength distribution and initial choice of estimation method. In general, one would expect a bootstrap method of this nature to perform best when the strength parameters from which the bootstrap is simulated are as similar to the actual values as possible, which would be supportive of using the $\alpha$-adjustment for the initial estimates under adaptive schemes, based on the evidence of the previous sections. 

\begin{figure}[hbtp]
    \centering
    \includegraphics[width=16cm]{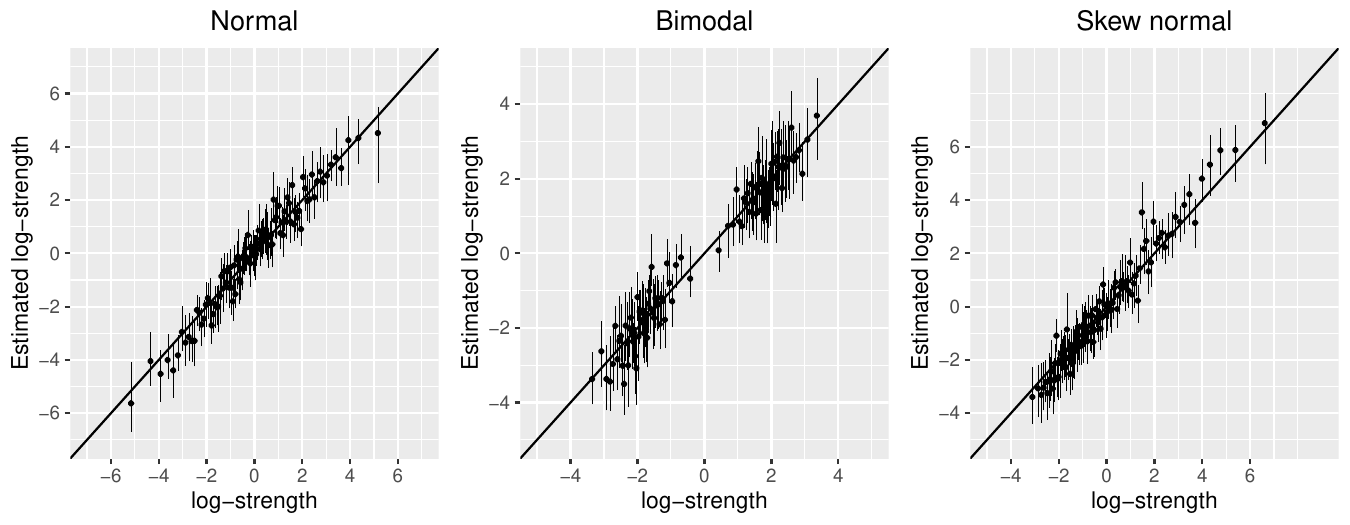}
    \caption{Caterpillar plot illustrating the sample 95\% confidence intervals obtained as a byproduct of the bootstrap bias correction for a Swiss scheduling scheme tournament of 20 rounds.}
    \label{fig: bias_corrected caterpillar}
\end{figure}

It should also be noted that this bootstrap method can be parallelised and so with modern computing can be run efficiently and relatively fast, even given large data sets. Another advantage of the method is that the simulations used in the bias correction calculation can be used to evaluate uncertainty surrounding the parameter estimates, which would otherwise have to come from asypmtotic assumptions that are unreliable in low data settings. An example is given in Figure \ref{fig: bias_corrected caterpillar}, where the 97.5\textsuperscript{th} and 2.5\textsuperscript{th} percentile ${}_s\lambda_i$ are used for each item $i$ in place of the term $\frac{1}{m}\sum^m_{s=1} {}_s\lambda_i$ in equation (\ref{eqn: bias correction}) to create a 95\% confidence interval for the estimate. The proportion of these where the true value lies within the confidence interval is suggestive of the accuracy of these error bounds.

\section{Concluding remarks}\label{sn: concluding}
The work presented here provides recommendations for current practice in CJ and suggestions for future research. Recommendations include:
\begin{enumerate}
    \item CJ analysts should give greater consideration to their parameter estimation methods. Authors should be more explicit about these choices and to publish code and appropriately anonymised comparison data along with their work. Without knowing what estimation procedure was used, it is not possible to adequately verify or challenge conclusions. Publishing data would allow later researchers to apply proposed methods to real data sets, accelerating the development of better practice. 
    \item For random sampling schemes, the penalization proposed by \citet{firth1993bias} should be used.  While all methods performed well on the random data here, $\epsilon$-adjustment, $\alpha$-adjustment, and dummy-adjustment all rely on a constant, the value of which is not strongly suggested by theory. On the other hand, the penalization method of \citet{firth1993bias} is free from such arbitrariness and has appealing asymptotic qualities \citep{firth1993bias,kosmidis2009bias, kosmidis2014bias}.
    \item For adaptive scheduling schemes, where possible, parameter estimation should be performed using a bias-corrected method based on an initial estimation using an $\alpha$-adjustment penalty. Work here suggests that a value of between $\alpha=0.3$ and $\alpha=0.5$ may be appropriate, though the data here suggest that the bootstrap estimation is robust to the initial estimation method.
    \item In circumstances where data are being analysed that have been produced through adaptive scheduling but the exact specification of that scheduling is not available, then it is recommended to use the $\alpha$-adjustment penalty. A value of $\alpha=0.5$ is tentatively suggested based on the results in Section \ref{sn: empirical}.
\end{enumerate}

Further research work is required around parameter estimation in CJ. Given the sparse data typical of CJ assessments, it is likely that bootstrap methods will remain appealing for both point and error estimation. But there is potential for further investigation as to how sparse those data can be. Additionally, performant adaptive methods are likely to depend on the estimation of item strength for online scheduling and for that task bootstrap methods may be limiting given their computational expense. Further, a drawback of the bias-corrected method is that full details of the scheduling scheme are required for its implementation and it may not be accessible to some practitioners seeking to use CJ unless it is implemented within the platform where they perform the scheduling. In Section \ref{sn: inference}, we proposed consideration of a pseudo-likelihood because of the difficulties of estimating $\mathbb{P}(C \mid M ; \boldsymbol{\lambda})$ and $\mathbb{P}(M ; \boldsymbol{\lambda})$, but perhaps under certain conditions these can be well-approximated allowing a penalization approach, without the requirement for using a bootstrap. Alternatively, perhaps good approximations to the expected information matrix can be formulated, which could then be applied through the method proposed by \citet{firth1993bias}.

With this in mind, it is worth noting that population sizes in CJ assessments are distributed bimodally. Many assessments are small with the number of items in the tens or low hundreds in line with the typical number of students in a school or university class or cohort. On the other hand, the national assessments conducted by No More Marking include more than fifty thousand scripts \citep{wheadon2020comparative}. Appropriate answers to some of the further research questions raised here may be dependent on population size and to what degree it is reasonable to make pre-analysis assumptions about the shape of the distribution of item log-strengths.

\bibliography{bibliography.bib}

\begin{thebibliography}{}

\bibitem[Andrich, 1978]{andrich1978relationships}
Andrich, D. (1978).
\newblock Relationships between the {T}hurstone and {R}asch approaches to item scaling.
\newblock {\em Applied Psychological Measurement}, 2(3):451--462.

\bibitem[Barlow et~al., 2020]{barlow2020inference}
Barlow, A.~M., Sherlock, C., and Tawn, J. (2020).
\newblock Inference for extreme values under threshold-based stopping rules.
\newblock {\em Journal of the Royal Statistical Society Series C: Applied Statistics}, 69(4):765--789.

\bibitem[Barrada et~al., 2008]{barrada2008incorporating}
Barrada, J.~R., Olea, J., Ponsoda, V., and Abad, F.~J. (2008).
\newblock Incorporating randomness in the {F}isher information for improving item-exposure control in {CAT}s.
\newblock {\em British Journal of Mathematical and Statistical Psychology}, 61(2):493--513.

\bibitem[Barrada et~al., 2010]{barrada2010method}
Barrada, J.~R., Olea, J., Ponsoda, V., and Abad, F.~J. (2010).
\newblock A method for the comparison of item selection rules in computerized adaptive testing.
\newblock {\em Applied Psychological Measurement}, 34(6):438--452.

\bibitem[Bartholomew and Jones, 2021]{bartholomew2021systematized}
Bartholomew, S.~R. and Jones, M.~D. (2021).
\newblock A systematized review of research with {A}daptive {C}omparative {J}udgment ({ACJ}) in higher education.
\newblock {\em International Journal of Technology and Design Education}, pages 1--32.

\bibitem[Bartholomew and Yoshikawa, 2018]{bartholomew2018systematic}
Bartholomew, S.~R. and Yoshikawa, E. (2018).
\newblock A systematic review of research around {A}daptive {C}omparative {J}udgment ({ACJ}) in {K}-16 education.
\newblock {\em Council on Technology an Engineering Teacher Education: Research Monograph Series}.

\bibitem[Bertoli-Barsotti et~al., 2014]{bertoli2014estimating}
Bertoli-Barsotti, L., Lando, T., and Punzo, A. (2014).
\newblock Estimating a {R}asch model via fuzzy empirical probability functions.
\newblock In {\em Analysis and modeling of complex data in behavioral and social sciences}, pages 29--36. Springer.

\bibitem[Bradley and Terry, 1952]{bradley1952rank}
Bradley, R.~A. and Terry, M.~E. (1952).
\newblock Rank analysis of incomplete block designs: {I}. {T}he method of paired comparisons.
\newblock {\em Biometrika}, 39(3/4):324--345.

\bibitem[Bramley, 2007]{bramley2007paired}
Bramley, T. (2007).
\newblock Paired comparison methods.
\newblock In Newton, P., Baird, J.-A., Goldstein, H., Patrick, H., and Tymms, P., editors, {\em Techniques for monitoring the comparability of examination standards}, pages 246--294. London: Qualifications and Curriculum Authority.

\bibitem[Bramley, 2015]{bramley2015investigating}
Bramley, T. (2015).
\newblock Investigating the reliability of adaptive comparative judgment.
\newblock {\em Cambridge Assessment, Cambridge}, 36.

\bibitem[Bramley and Vitello, 2019]{bramley2019effect}
Bramley, T. and Vitello, S. (2019).
\newblock The effect of adaptivity on the reliability coefficient in adaptive comparative judgement.
\newblock {\em Assessment in Education: Principles, Policy \& Practice}, 26(1):43--58.

\bibitem[B\"{u}hlmann and Huber, 1963]{buhlmann1963pairwise}
B\"{u}hlmann, H. and Huber, P.~J. (1963).
\newblock Pairwise comparison and ranking in tournaments.
\newblock {\em The Annals of Mathematical Statistics}, 34(2):501--510.

\bibitem[Chen and Smith, 1984]{chen1984bayes}
Chen, C. and Smith, T.~M. (1984).
\newblock A {B}ayes-type estimator for the {B}radley-{T}erry model for paired comparison.
\newblock {\em Journal of statistical planning and inference}, 10(1):9--14.

\bibitem[Cox, 1958]{cox1958some}
Cox, D.~R. (1958).
\newblock Some problems connected with statistical inference.
\newblock {\em The Annals of Mathematical Statistics}, 29(2):357--372.

\bibitem[Crompvoets et~al., 2020]{crompvoets2020adaptive}
Crompvoets, E.~A., B{\'e}guin, A.~A., and Sijtsma, K. (2020).
\newblock Adaptive pairwise comparison for educational measurement.
\newblock {\em Journal of Educational and Behavioral Statistics}, 45(3):316--338.

\bibitem[Davidson and Solomon, 1973]{davidson1973bayesian}
Davidson, R.~R. and Solomon, D.~L. (1973).
\newblock A {B}ayesian approach to paired comparison experimentation.
\newblock {\em Biometrika}, 60(3):477--487.

\bibitem[Davies et~al., 2020]{davies2020comparative}
Davies, B., Alcock, L., and Jones, I. (2020).
\newblock Comparative judgement, proof summaries and proof comprehension.
\newblock {\em Educational Studies in Mathematics}, 105(2):181--197.

\bibitem[Firth, 1993]{firth1993bias}
Firth, D. (1993).
\newblock Bias reduction of maximum likelihood estimates.
\newblock {\em Biometrika}, 80(1):27--38.

\bibitem[Glickman and Jensen, 2005]{glickman2005adaptive}
Glickman, M.~E. and Jensen, S.~T. (2005).
\newblock Adaptive paired comparison design.
\newblock {\em Journal of Statistical Planning and Inference}, 127(1-2):279--293.

\bibitem[Goffin and Olson, 2011]{goffin2011all}
Goffin, R.~D. and Olson, J.~M. (2011).
\newblock Is it all relative? {C}omparative judgments and the possible improvement of self-ratings and ratings of others.
\newblock {\em Perspectives on Psychological Science}, 6(1):48--60.

\bibitem[Haberman, 2004]{haberman2004joint}
Haberman, S.~J. (2004).
\newblock Joint and conditional maximum likelihood estimation for the {R}asch model for binary responses.
\newblock {\em ETS Research Report Series}, 2004(1):i--63.

\bibitem[Jeffreys, 1946]{jeffreys1946invariant}
Jeffreys, H. (1946).
\newblock An invariant form for the prior probability in estimation problems.
\newblock {\em Proceedings of the Royal Society of London. Series A. Mathematical and Physical Sciences}, 186(1007):453--461.

\bibitem[Jones and Alcock, 2012]{jones2012summative}
Jones, I. and Alcock, L. (2012).
\newblock Summative peer assessment of undergraduate calculus using adaptive comparative judgement.
\newblock {\em Mapping University Mathematics Assessment Practices}, pages 63--74.

\bibitem[Jones and Davies, 2023]{jones2023comparative}
Jones, I. and Davies, B. (2023).
\newblock Comparative judgement in education research.
\newblock {\em International Journal of Research \& Method in Education}, pages 1--12.

\bibitem[Jones and Sirl, 2017]{jones2017peer}
Jones, I. and Sirl, D. (2017).
\newblock Peer assessment of mathematical understanding using comparative judgement.
\newblock {\em Nordic Studies in Mathematics Education}, 22(4).

\bibitem[Kenne~Pagui et~al., 2017]{kenne2017median}
Kenne~Pagui, E.~C., Salvan, A., and Sartori, N. (2017).
\newblock Median bias reduction of maximum likelihood estimates.
\newblock {\em Biometrika}, 104(4):923--938.

\bibitem[Kosmidis, 2014]{kosmidis2014bias}
Kosmidis, I. (2014).
\newblock Bias in parametric estimation: reduction and useful side-effects.
\newblock {\em Wiley Interdisciplinary Reviews: Computational Statistics}, 6(3):185--196.

\bibitem[Kosmidis, 2020]{kosmidis2020brglm2}
Kosmidis, I. (2020).
\newblock brglm2: Bias reduction in generalized linear models.
\newblock {\em R package version 0.6}, 2:635.

\bibitem[Kosmidis and Firth, 2009]{kosmidis2009bias}
Kosmidis, I. and Firth, D. (2009).
\newblock Bias reduction in exponential family nonlinear models.
\newblock {\em Biometrika}, 96(4):793--804.

\bibitem[Kosmidis and Firth, 2011]{kosmidis2011multinomial}
Kosmidis, I. and Firth, D. (2011).
\newblock Multinomial logit bias reduction via the {P}oisson log-linear model.
\newblock {\em Biometrika}, 98(3):755--759.

\bibitem[Kosmidis and Firth, 2021]{kosmidis2021jeffreys}
Kosmidis, I. and Firth, D. (2021).
\newblock Jeffreys-prior penalty, finiteness and shrinkage in binomial-response generalized linear models.
\newblock {\em Biometrika}, 108(1):71--82.

\bibitem[Kosmidis et~al., 2020]{kosmidis2020mean}
Kosmidis, I., Kenne~Pagui, E.~C., and Sartori, N. (2020).
\newblock Mean and median bias reduction in generalized linear models.
\newblock {\em Statistics and Computing}, 30(1):43--59.

\bibitem[Laming, 2003]{laming2003human}
Laming, D. (2003).
\newblock {\em Human judgment: The eye of the beholder}.
\newblock Cengage Learning EMEA.

\bibitem[Leonard, 1977]{leonard1977alternative}
Leonard, T. (1977).
\newblock An alternative {B}ayesian approach to the {B}radley-{T}erry model for paired comparisons.
\newblock {\em Biometrics}, pages 121--132.

\bibitem[Linacre, 2004]{linacre2004rasch}
Linacre, J.~M. (2004).
\newblock Rasch model estimation: Further topics.
\newblock {\em Journal of Applied Measurement}, 5(1):95--110.

\bibitem[McCullagh and Nelder, 1989]{mccullagh1989generalized}
McCullagh, P. and Nelder, J.~A. (1989).
\newblock {\em Generalized linear models}.
\newblock Chapman and Hall.

\bibitem[Molenaar, 1995]{molenaar1995estimation}
Molenaar, I.~W. (1995).
\newblock Estimation of item parameters.
\newblock In {\em Rasch Models}, pages 39--51. Springer.

\bibitem[Pfeiffer et~al., 2012]{pfeiffer2012adaptive}
Pfeiffer, T., Gao, X., Chen, Y., Mao, A., and Rand, D. (2012).
\newblock Adaptive polling for information aggregation.
\newblock In {\em Proceedings of the AAAI Conference on Artificial Intelligence}, volume~26.

\bibitem[Phelan and Whelan, 2017]{phelan2017hierarchical}
Phelan, G.~C. and Whelan, J.~T. (2017).
\newblock Hierarchical {B}ayesian {B}radley-{T}erry for applications in major league baseball.
\newblock {\em arXiv preprint arXiv:1712.05879}.

\bibitem[Pinot~de Moira et~al., 2022]{pinot2022classification}
Pinot~de Moira, A., Wheadon, C., and Christodoulou, D. (2022).
\newblock The classification accuracy and consistency of comparative judgement of writing compared to rubric-based teacher assessment.
\newblock {\em Research in Education}, 113(1):25--40.

\bibitem[Pollitt, 2012a]{pollitt2012comparative}
Pollitt, A. (2012a).
\newblock Comparative judgement for assessment.
\newblock {\em International Journal of Technology and Design Education}, 22(2):157--170.

\bibitem[Pollitt, 2012b]{pollitt2012method}
Pollitt, A. (2012b).
\newblock The method of adaptive comparative judgement.
\newblock {\em Assessment in Education: principles, policy \& practice}, 19(3):281--300.

\bibitem[Pollitt, 2015]{pollitt2015reliability}
Pollitt, A. (2015).
\newblock On “reliability” bias in {ACJ}: Valid simulation of adaptive comparative judgement.
\newblock {\em Cambridge Exam Research, Cambridge, England}.

\bibitem[Quenouille, 1949]{quenouille1949approximate}
Quenouille, M.~H. (1949).
\newblock Approximate tests of correlation in time-series.
\newblock {\em Journal of the Royal Statistical Society: Series B}, 11:68--84.

\bibitem[Quenouille, 1956]{quenouille1956notes}
Quenouille, M.~H. (1956).
\newblock Notes on bias in estimation.
\newblock {\em Biometrika}, 43(3/4):353--360.

\bibitem[{R Core Team}, 2021]{RCore2021alanguage}
{R Core Team} (2021).
\newblock {\em R: {A} Language and Environment for Statistical Computing}.
\newblock R Foundation for Statistical Computing, Vienna, Austria.

\bibitem[Rangel-Smith and Lynch, 2018]{rangel2018addressing}
Rangel-Smith, C. and Lynch, D. (2018).
\newblock Addressing the issue of bias in the measurement of reliability in the method of adaptive comparative judgment.
\newblock In {\em PATT36 international conference. Research \& Practice in Technology Education: Perspectives on Human Capacity and Development}, pages 378--388.

\bibitem[Revuelta and Ponsoda, 1998]{revuelta1998comparison}
Revuelta, J. and Ponsoda, V. (1998).
\newblock A comparison of item exposure control methods in computerized adaptive testing.
\newblock {\em Journal of Educational Measurement}, 35(4):311--327.

\bibitem[Robitzsch, 2021]{robitzsch2021comprehensive}
Robitzsch, A. (2021).
\newblock A comprehensive simulation study of estimation methods for the {R}asch model.
\newblock {\em Stats}, 4(4):814--836.

\bibitem[Robitzsch and Robitzsch, 2022]{robitzsch2022package}
Robitzsch, A. and Robitzsch, M.~A. (2022).
\newblock Package ‘sirt’.

\bibitem[Warm, 1989]{warm1989weighted}
Warm, T.~A. (1989).
\newblock Weighted likelihood estimation of ability in item response theory.
\newblock {\em Psychometrika}, 54(3):427--450.

\bibitem[Wheadon, 2015a]{wheadon2015analysing}
Wheadon, C. (2015a).
\newblock {\em Analysing comparative judgement data in {R}}.
\newblock \url{https://blog.nomoremarking.com/analysing-comparative-judgement-data-in-r-19ac5924602a}, accessed June 7, 2022.

\bibitem[Wheadon, 2015b]{wheadon2015opposite}
Wheadon, C. (2015b).
\newblock The opposite of adaptivity?
\newblock \url{https://blog.nomoremarking.com/the-opposite-of-adaptivity-c26771d21d50}, accessed September 27, 2022.

\bibitem[Wheadon et~al., 2020]{wheadon2020comparative}
Wheadon, C., Barmby, P., Christodoulou, D., and Henderson, B. (2020).
\newblock A comparative judgement approach to the large-scale assessment of primary writing in {E}ngland.
\newblock {\em Assessment in Education: Principles, Policy \& Practice}, 27(1):46--64.

\bibitem[Whelan, 2017]{whelan2017prior}
Whelan, J.~T. (2017).
\newblock Prior distributions for the {B}radley-{T}erry model of paired comparisons.
\newblock {\em arXiv preprint arXiv:1712.05311}.

\bibitem[Wobus, 2007]{wobus2007KRACH}
Wobus, J. (2007).
\newblock Krach ratings.
\newblock \url{http://sports.vaporia.com/krach.html}, accessed October 4, 2022.

\bibitem[Zermelo, 1929]{zermelo1929berechnung}
Zermelo, E. (1929).
\newblock Die {B}erechnung der {T}urnier-ergebnisse als ein {M}aximumproblem der {W}ahrscheinlichkeitsrechnung.
\newblock {\em Mathematische Zeitschrift}, 29(1):436--460.

\end{thebibliography}

\end{document}